\pdfoutput=1

\documentclass[11pt,twoside,a4paper,cmspaper,final,collab]{cms-tdr}
\def\svnVersion{195901}\def\cmsCernNo{2013-077}\def\cmsCernDate{\today}\def\cmsMessage{Published in Physics Letters B as \href{http://dx.doi.org/10.1016/j.physletb.2013.06.028}{\doi{10.1016/j.physletb.2013.06.028.}}}

\begin{document}\cmsNoteHeader{HIN-13-002}

\hyphenation{had-ron-i-za-tion}
\hyphenation{cal-or-i-me-ter}
\hyphenation{de-vices}

\RCS$Revision: 184410 $
\RCS$HeadURL: svn+ssh://svn.cern.ch/reps/tdr2/papers/HIN-13-002/trunk/HIN-13-002.tex $
\RCS$Id: HIN-13-002.tex 184410 2013-05-03 01:52:07Z alverson $
\newlength\cmsFigWidth
\ifthenelse{\boolean{cms@external}}{\setlength\cmsFigWidth{0.98\columnwidth}}{\setlength\cmsFigWidth{0.7\textwidth}}
\ifthenelse{\boolean{cms@external}}{\providecommand{\cmsLeft}{top}}{\providecommand{\cmsLeft}{left}}
\ifthenelse{\boolean{cms@external}}{\providecommand{\cmsRight}{bottom}}{\providecommand{\cmsRight}{right}}
\providecommand{\re}{\ensuremath{\cmsSymbolFace{e}}}
\newcommand {\roots}    {\ensuremath{\sqrt{s}}}
\newcommand {\rootsNN}  {\ensuremath{\sqrt{s_{_{NN}}}}}

\newcommand {\pp}    {\mbox{pp}}
\newcommand {\ppbar} {\mbox{p\={p}}}
\newcommand {\pbarp} {\mbox{p\={p}}}
\newcommand {\Pb}    {\mbox{Pb}}
\newcommand {\PbPb}  {\mbox{PbPb}}
\newcommand {\AuAu}  {\mbox{AuAu}}
\newcommand {\pPb}  {\ensuremath{\text{pPb}}\xspace}
\newcommand {\AonA}  {\ensuremath{\text{AA}}\xspace}
\newcommand {\pA}    {\ensuremath{\text{pA}}\xspace}
\newcommand{\noff}    {\ensuremath{N_\text{trk}^\text{offline}}\xspace}
\newcommand{\npri}    {\ensuremath{N_\text{trk}^\text{best}}\xspace}
\newcommand{\nsec}    {\ensuremath{N_\text{trk}^\text{add}}\xspace}
\newcommand {\photonjet}{photon+jet}
\newcommand {\pttrg}       {\ensuremath{p_\mathrm{T}^{\text{trig}}}}
\newcommand {\ptass}       {\ensuremath{p_\mathrm{T}^{\text{assoc}}}}
\newcommand {\ptref}       {\ensuremath{p_\mathrm{T}^{\text{ref}}}}
\newcommand {\deta}     {\ensuremath{\Delta\eta}}
\newcommand {\dphi}     {\ensuremath{\Delta\phi}}
\providecommand{\EPOS} {\textsc{epos}\xspace}

\cmsNoteHeader{HIN-13-002} 
\title{Multiplicity and transverse-momentum dependence of two- and four-particle correlations in \pPb\ and \PbPb\ collisions}

\date{\today}

\abstract{
Measurements of two- and four-particle angular correlations for charged
particles emitted in \pPb\ collisions are presented over a wide range in
pseudorapidity and full azimuth. The data, corresponding to an integrated
luminosity of approximately 31\nbinv, were collected during the 2013
LHC \pPb\ run at a nucleon-nucleon center-of-mass energy of 5.02\TeV by the CMS experiment.
The results are compared to 2.76\TeV
semi-peripheral \PbPb\ collision data, collected during the 2011 \PbPb\ run,
covering a similar range of particle multiplicities.
The observed correlations are characterized by the near-side ($\abs{\dphi} \approx 0$) associated
pair yields and the azimuthal anisotropy Fourier harmonics ($v_n$).
The second-order ($v_2$) and third-order ($v_3$) anisotropy harmonics are
extracted using the two-particle azimuthal correlation technique. A four-particle
correlation method is also applied to obtain the value of $v_2$
and further explore the multi-particle nature of the correlations.
Both associated pair yields and anisotropy harmonics are
studied as a function of particle multiplicity and transverse momentum.
The associated pair yields, the four-particle $v_2$, and the
$v_3$ become apparent at about the same multiplicity.
A remarkable similarity in the $v_3$ signal as a function of multiplicity is observed between the \pPb\ and \PbPb\ systems.
Predictions based on the color glass condensate and hydrodynamic
models are compared to the experimental results.
}

\hypersetup{%
pdfauthor={CMS Collaboration},%
pdftitle={Multiplicity and transverse-momentum dependence of two- and four-particle correlations in pPb and PbPb collisions},%
pdfsubject={CMS},%
pdfkeywords={CMS, physics, heavy ions}}

\maketitle 
\section{Introduction}
\label{sec:intro}

Studies of multi-particle correlations play a major role in
characterizing the underlying mechanism of particle production in high-energy
collisions of protons and nuclei. Of particular interest in
relativistic nucleus-nucleus (AA) collisions is the observed long-range
(large $|\deta|$) structure in two-dimensional (2D) $\deta$-$\dphi$
correlation functions. Here, $\dphi$ and $\deta$ are the differences in
azimuthal angle $\phi$ and pseudorapidity $\eta$ =$-\ln[\tan(\theta/2)]$
between the two particles, where the polar angle $\theta$ is defined relative
to the beam axis. One source of such long-range correlations
is the ``elliptic flow'' induced by the hydrodynamic evolution of
the lenticular collision zone
in non-central nucleus-nucleus interactions~\cite{Ollitrault:1992bk}.
Elliptic flow contributes a $\cos(2\dphi)$ component to the two-particle
correlation function over a broad $|\deta|$ range~\cite{Alver:2008gk}.
The studies of elliptic flow have been carried out over a wide
range of energies and collision systems~\cite{Adcox:2004mh,Adams:2005dq,Back:2004je,Adler:2004cj,
Agakishiev:2011id,PHOBOS:PhysRevLett.98.242302,
Chatrchyan:2012ta,ALICE:PhysRevLett.105.252302,ATLAS:2011ah}.

After subtracting the elliptic flow component, a pronounced correlation
structure at $\abs{\dphi} \approx 0$ (near-side) extending over large $|\deta|$ remains
in \AuAu\ collisions at the Relativistic Heavy Ion Collider
(RHIC)~\cite{Adams:2005ph,Abelev:2009af,Alver:2009id,Abelev:2009jv}.
Such long-range near-side correlations are not reproduced by models of
nucleon-nucleon interactions such as \PYTHIA, and are not observed in \pp\ collisions at RHIC energies.
A variety of theoretical models were proposed to interpret this
residual long-range near-side correlation as a consequence of jet-medium
interactions~\cite{Armesto:2004pt,Majumder:2006wi,Chiu:2005ad,Wong:2008yh,Romatschke:2006bb,Shuryak:2007fu}.
However, it was later realized that, because of event-by-event fluctuations in the
initial-state collision geometry~\cite{Voloshin:2004th,Mishra:2007tw,Takahashi:2009na},
higher-order anisotropic flow components could also be
induced, in particular the ``triangular flow'' which contributes a $\cos(3\dphi)$ component
~\cite{Alver:2010gr,Alver:2010dn,Schenke:2010rr,
Petersen:2010cw,Xu:2010du,Teaney:2010vd}. Therefore, the observed long-range $\deta$
correlations in \AonA\ collisions can, in general, be attributed to the collective
expansion of a strongly-interacting medium. The detailed azimuthal correlation
function is typically characterized by its Fourier components,
$\sim 1+2 \sum_{n}^{} v_{n}^{2}\cos(n\dphi)$, where $v_n$ denote
the single-particle anisotropy harmonics~\cite{Voloshin:1994mz}. In particular,
the second (elliptic) and third (triangular) Fourier harmonics are assumed to
most directly reflect the medium response to the initial collision geometry
and to its fluctuations, respectively. Detailed studies of elliptic and
triangular flow provide insight into fundamental transport properties of the medium~\cite{Alver:2010dn,Schenke:2010rr,Qiu:2011hf}. The long-range correlations
and anisotropy Fourier harmonics have also been extensively studied in \PbPb\
collisions at the Large Hadron Collider (LHC)~\cite{Chatrchyan:2011eka,Chatrchyan:2012wg,Chatrchyan:2012ta,ALICE:PhysRevLett.105.252302,ALICE:2011ab,Aamodt:2011by,ATLAS:2011ah,ATLAS:2012at}.

Recently, a similar long-range near-side correlation structure (without subtraction of any flow component)
was observed in the highest particle multiplicity proton-proton (\pp)~\cite{Khachatryan:2010gv}
and proton-lead (\pPb)~\cite{CMS:2012qk} collisions at the LHC.
In \pPb\ collisions, the overall strength
of the near-side correlation is found to be significantly greater than in \pp\ collisions.
The away-side ($\abs{\dphi} \approx \pi$) correlations contain substantial contributions
from the back-to-back jets, and thus have not been the focus of these initial studies.
A procedure for removing the jet correlations on the away
side by subtracting the correlations for very-low-multiplicity data was
recently introduced~\cite{alice:2012qe,atlas:2012fa}, and used to study the long-range correlations in \pPb\
on both near and away sides using the anisotropy Fourier harmonics.
Evidence of such
correlations was also found recently in 200\GeV deuteron-gold collisions at
RHIC~\cite{Adare:2013piz}.
While hydrodynamic flow is the commonly accepted explanation
of such long-range correlations in the \AonA\ collision systems, a variety of theoretical models
have been proposed to explain the origin of this phenomenon in small collision
systems like \pp\ (see Ref.~\cite{Li:2012hc} for a recent review) and \pPb.
Such models include gluon saturation in the initial interaction of the
protons and nuclei~\cite{Dusling:2012cg,Dusling:2012wy} and hydrodynamic effects in the
high-density systems possibly formed in these collisions at \TeV
energies~\cite{Bozek:2011if,Bozek:2012gr}. Since hydrodynamic flow is intrinsically
a multi-particle phenomenon, it can be probed more directly using multi-particle correlation
(or cumulant) techniques~\cite{Bilandzic:2010jr} rather than with two-particle
correlations. In particular,
two-particle correlations, arising from jet production, are expected to
be strongly suppressed using the multi-particle method. A measurement of an elliptic
flow signal using the four-particle cumulant method in \pPb\ collisions was recently
presented~\cite{Aad:2013fja}.

To provide further constraints on the theoretical understanding of the
particle production mechanisms in different collision systems, this paper
presents a detailed analysis of two- and four-particle angular correlations
in \pPb\ collisions at \rootsNN\ = 5.02\TeV. This 2013 data set, especially with
the implementation of a dedicated high-multiplicity trigger, provides a much larger
sample of very-high-multiplicity \pPb\ events. Therefore, correlations can be
explored up to a multiplicity comparable to that in mid-central \PbPb\ collisions
(e.g., $\sim$55\% centrality, where centrality is defined as
the fraction of the total inelastic cross section, with 0\% denoting the most
central collisions). The two-particle long-range correlation
data are presented in two different but closely-related
approaches: the near-side associated yields, which characterize the absolute yield
of correlated particle pairs, and anisotropy harmonics ($v_2$ and $v_3$), which provide
a measurement of relative correlation magnitude with respect to the uncorrelated
background. To further investigate
the multi-particle nature of the correlations, a four-particle cumulant analysis
is also performed for determining the $v_2$ harmonic. Both the associated yields
and anisotropy harmonics are studied as a function of particle multiplicity
and transverse momentum, providing a direct comparison of \pPb\ and \PbPb\
collision systems over a broad range of similar multiplicities.

\section{Experimental Setup}
\label{sec:exp_evt}

The CMS detector comprises a number of subsystems and a detailed description
can be found in Ref.~\cite{JINST}. The results in this paper are mainly based on the
silicon tracker information. This detector, located in the 3.8\unit{T} field of the
superconducting solenoid, consists of 1\,440 silicon pixel and 15\,148 silicon strip
detector modules. The silicon tracker measures charged particles within the pseudorapidity
range $\abs{\eta}< 2.5$, and provides an impact parameter resolution of ${\approx}15\mum$ and
a transverse momentum (\pt) resolution better than 1.5\% up to $\pt \approx 100\GeVc$.
The electromagnetic calorimeter (ECAL) and hadron calorimeter (HCAL) are also
located inside the solenoid. The ECAL consists of 75\,848
lead-tungstate crystals, arranged in a quasi-projective geometry and distributed in a
barrel region ($\abs{\eta} < 1.48$) and two endcaps that extend to $\abs{\eta} = 3.0$.
The HCAL barrel and endcaps are sampling calorimeters composed of brass and
scintillator plates, covering $\abs{\eta} < 3.0$. Iron/quartz-fiber \v Cerenkov
hadron forward (HF) calorimeters cover the range $2.9 < \abs{\eta} < 5.2$ on either
side of the interaction region.
The detailed Monte Carlo (MC) simulation of the CMS detector response is based
on \GEANTfour~\cite{GEANT4}.

\section{Selections of Events and Tracks}

This analysis is performed using data recorded by CMS during the
LHC \pPb\ run in 2013. The data set corresponds
to an integrated luminosity of about 31\nbinv, assuming a pPb interaction
cross section of 2.1\unit{barns}. The beam energies were 4\TeV for protons and 1.58\TeV
per nucleon for lead nuclei, resulting in a center-of-mass energy per nucleon pair
of 5.02\TeV. The direction of the higher energy proton beam was initially
set up to be clockwise, and was then reversed.
As a result of the energy difference between the colliding beams, the nucleon-nucleon
center-of-mass in the \pPb\ collisions is not at rest with respect to the laboratory frame.
Massless particles emitted at $\eta_\unit{cm} = 0$ in the nucleon-nucleon
center-of-mass frame will be detected at $\eta = -0.465$ (clockwise proton beam)
or $0.465$ (counterclockwise proton beam) in the laboratory frame.
A sample of 2.76\TeV \PbPb\ data collected during the 2011 LHC heavy-ion run,
corresponding to an integrated luminosity of 2.3\mubinv, is also analyzed for
comparison purposes.

Minimum bias (MB) \pPb\ events were triggered by requiring at least
one track with $\pt > 0.4$\GeVc to be found in the pixel tracker
for a \pPb\ bunch crossing. Because of hardware limits on the data
acquisition rate, only a small fraction (${\sim}10^{-3}$) of all minimum bias
triggered events were recorded (i.e., the trigger was ``prescaled'').
In order to select high-multiplicity \pPb\ collisions,
a dedicated high-multiplicity trigger was implemented using the CMS level-1
(L1) and high-level trigger (HLT) systems. At L1, the total transverse energy
summed over ECAL and HCAL was required to be greater than a given threshold
(20 or 40\GeV). Online track reconstruction for the HLT was based on
the three layers of pixel detectors, and required a track origin within a cylindrical region of
length 30\unit{cm} along the beam and radius 0.2\unit{cm} perpendicular to the beam.
For each event, the vertex reconstructed with the highest number of pixel tracks was selected.
The number of pixel tracks (${N}_\text{trk}^\text{online}$)
with $\abs{\eta}<2.4$, $\pt > 0.4\GeVc$, and a distance of closest approach of 0.4\unit{cm} or
less to this vertex, was determined for each event. Data were taken with thresholds of
${N}_\text{trk}^\text{online}>100, 130$ (L1 threshold of 20\GeV), and $160, 190$
(L1 threshold of 40\GeV) with prescaling factors dependent on the instantaneous luminosity.
The ${N}_\text{trk}^\text{online}>190$ trigger was never prescaled throughout
the entire run.

In the offline analysis, hadronic collisions were selected by requiring a
coincidence of at least one HF calorimeter tower with more than 3\GeV of
total energy in each of the HF detectors.
Events were also required to contain at least one reconstructed primary
vertex within 15\unit{cm} of the nominal interaction point along the beam axis
and within 0.15\unit{cm} transverse to the beam trajectory.
At least two reconstructed tracks were required to be associated
with the primary vertex. Beam related background was suppressed
by rejecting events for which less than 25\% of all reconstructed tracks were
of good quality (i.e., the tracks selected for physics analysis as will be discussed later).

The \pPb\ instantaneous luminosity provided by the LHC in the 2013
run resulted in approximately 3\% probability of at least one additional interaction
occurring in the same bunch crossing, resulting in pileup events. A procedure for rejecting pileup
events was developed to select clean, single-vertex \pPb\ collisions.
The approach was to investigate the number of tracks, \npri\ that is assigned
to the best reconstructed vertex (e.g., the one with the greatest number of
associated tracks), and \nsec\ assigned to each of the additional vertices,
as well as the distance between the two vertices in the $z$ direction ($\Delta z_\text{vtx}$).
Based on studies using low pileup \pPb\ data (from the 2012 pilot run),
\PbPb\ data, and MC simulations, events with \nsec\ above a certain threshold at a
given $\Delta z_\text{vtx}$ were identified as pileup events and removed from the event sample.
This threshold was set to be higher for smaller $\Delta z_\text{vtx}$ and larger \npri\ to account
for the fact that events with a smaller vertex separation and greater multiplicity
have a higher probability of vertex splitting in the reconstruction algorithm.
The residual pileup fraction was estimated to be no more than 0.2\% for the highest
multiplicity \pPb\ interactions studied in this paper.

Among those \pPb\ interactions simulated with the \EPOS~\cite{Porteboeuf:2010um}
and \HIJING~\cite{Gyulassy:1994ew} event generators, which have at least one primary
particle with total energy $E>3$\GeV in both $\eta$ ranges of $-5<\eta<-3$ and
$3<\eta<5$, the above criteria are found to select 97--98\% of the events.

In this analysis, the CMS \textit{highPurity}~\cite{Stenson:2010xx} tracks
were used. Additionally, a reconstructed track was only considered as a
primary-track candidate if the significance of the separation along the beam axis ($z$)
between the track and the best vertex, $d_z/\sigma(d_z)$, and the significance
of the impact parameter relative to the best vertex transverse to the beam,
$d_\mathrm{T}/\sigma(d_\mathrm{T})$, were each less than 3. The relative uncertainty
of the transverse-momentum measurement, $\sigma(\pt)/\pt$, was required to be less than 10\%.
To ensure high tracking efficiency and reduce the rate of misidentified tracks,
only tracks within $\abs{\eta}<2.4$ and with $\pt >$ 0.3\GeVc were used in the analysis
(a different \pt\ cutoff of 0.4\GeVc used in multiplicity determination due to constraint
of online processing time at HLT).

The events were divided into classes of reconstructed track multiplicity, \noff,
where primary tracks with $\abs{\eta}<2.4$ and $\pt >0.4$\GeVc were counted, in a method similar
to the approach used in Refs.~\cite{CMS:2012qk, Khachatryan:2010gv}. Data from the
HLT minimum bias trigger were used for $\noff<120$, while the track multiplicity triggers
with online track thresholds of 100, 130, 160, and 190 were used for $120 \leq \noff <150$,
$150 \leq \noff <185$, $185 \leq \noff <220$, and $\noff \geq 220$, respectively.
This correspondence ensures at least 90\% trigger efficiency in each multiplicity bin.
The fractions of MB triggered events after event selections falling into each of
the main multiplicity classes are listed in Table~\ref{tab:newmultbinning}.
The table also lists the average values of \noff\
and $N_\text{trk}^\text{corrected}$, the event
multiplicity of charged particles with $\abs{\eta}<2.4$ and $\pt >0.4$\GeVc corrected
for detector acceptance and efficiency of the track reconstruction algorithm,
as discussed in the following section. The average \noff\ values for MB \pPb\ samples
with opposite proton beam directions are found to be consistent within 0.2\%.

In order to compare directly the \pPb\ and \PbPb\ systems using event selections
based on the multiplicity of the collisions, a subset of data from peripheral \PbPb\
collisions collected during the 2011 LHC heavy-ion run with a minimum bias trigger
were reanalyzed using the same track reconstruction algorithm as the one employed
for \pp\ and \pPb\ collisions. The selection of events and tracks is the same as for the
present \pPb\ analysis although a different trigger is used.
A description of the 2011 \PbPb\ data can be found in
Ref.~\cite{Chatrchyan:2012xq}. The average \noff\ and $N_\text{trk}^\text{corrected}$ values,
and corresponding average \PbPb\ collision centrality, as determined by the total
energy deposited in the HF calorimeters~\cite{Chatrchyan:2012ta}, are listed in
Table~\ref{tab:newmultbinning} for each \noff\ bin.

\begin{table*}[ht]
\centering
\topcaption{\label{tab:newmultbinning} Fraction of MB triggered events after event selections
in each multiplicity bin, and the average multiplicity of reconstructed tracks per bin
with $\abs{\eta}<2.4$ and $\pt >0.4$\GeVc, before (\noff) and after ($N_\text{trk}^\text{corrected}$) efficiency
correction, for 2.76\TeV \PbPb\ and 5.02\TeV \pPb\ data.}
\begin{tabular}{ l | l  l  l | l  l  l }
\hline
\multicolumn{1}{l|}{} & \multicolumn{3}{c|}{\PbPb\ data} & \multicolumn{3}{c}{\pPb\ data}\\
\hline
\noff\ bin & $\langle \text{Centrality} \rangle$ & $\left<\noff \right>$ & $\left<N_\text{trk}^\text{corrected}\right>$ & Fraction & $\left<\noff \right>$ & $\left<N_\text{trk}^\text{corrected}\right>$ \\
& $\pm$ RMS (\%) & & & & & \\
\hline
$[0, \infty)$    & & & &   1.00  & 40 & 50$\pm$2 \\
$[0, 20)$    & 92$\pm$4   &  10   & 13$\pm$1 &   0.31  &  10   & 12$\pm$1      \\
$[20, 30)$   & 86$\pm$4   &  24   & 30$\pm$1 &   0.14  &  25   & 30$\pm$1      \\
$[30, 40)$   & 83$\pm$4   &  34   & 43$\pm$2 &   0.12  &  35   & 42$\pm$2      \\
$[40, 50)$   & 80$\pm$4   &  44   & 55$\pm$2 &   0.10  &  45   & 54$\pm$2      \\
$[50, 60)$   & 78$\pm$3   &  54   & 68$\pm$3 &   0.09  &  54   & 66$\pm$3      \\
$[60, 80)$   & 75$\pm$3   &  69   & 87$\pm$4 &   0.12  &  69   & 84$\pm$4      \\
$[80, 100)$  & 72$\pm$3   &  89   & 112$\pm$5 &   0.07  &  89   & 108$\pm$5    \\
$[100, 120)$ & 70$\pm$3   &  109  & 137$\pm$6 &   0.03  &  109  & 132$\pm$6    \\
$[120, 150)$ & 67$\pm$3   &  134  & 168$\pm$7 &   0.02  &  132  & 159$\pm$7    \\
$[150, 185)$ & 64$\pm$3   &   167  & 210$\pm$9 &   $4 \times 10^{-3}$   &  162  & 195$\pm$9     \\
$[185, 220)$ & 62$\pm$2   &  202  &  253$\pm$11 &   $5 \times 10^{-4}$   &  196  &  236$\pm$10  \\
$[220, 260)$ & 59$\pm$2   &  239  &  299$\pm$13 &   $6 \times 10^{-5}$   &  232  &  280$\pm$12  \\
$[260, 300)$ & 57$\pm$2   &  279  &  350$\pm$15 &   $3 \times 10^{-6}$   &  271  &  328$\pm$14  \\
$[300, 350)$ & 55$\pm$2   &  324  &  405$\pm$18 &   $1 \times 10^{-7}$   &  311  &  374$\pm$16  \\
\hline
\end{tabular}
\end{table*}

\section{Analysis Technique}
\label{sec:analysis}

\subsection{Two-Particle Correlation Function}
\label{subsec:analysis_dihadroncorr}

The two-particle correlation functions are constructed following
the procedure established in Refs.~\cite{Chatrchyan:2011eka,Chatrchyan:2012wg,CMS:2012qk}.
For each track multiplicity class, ``trigger'' particles are defined as
primary charged tracks within a given \pttrg\ range.
The number of trigger particles in the event is denoted by $N_\text{trig}$.
Particle pairs are formed by associating each trigger particle with
the remaining charged primary particles from a specified \ptass\ interval
(which can be either the same or different from the \pttrg\ range).
The per-trigger-particle associated yield
is defined as
\begin{equation}
\label{2pcorr_incl}
\frac{1}{N_\text{trig}}\frac{\rd^{2}N^\text{pair}}{\rd\Delta\eta\, \rd\Delta\phi}
= B(0,0)\times\frac{S(\Delta\eta,\Delta\phi)}{B(\Delta\eta,\Delta\phi)},
\end{equation}
where $\Delta\eta$ and $\Delta\phi$ are the differences in $\eta$
and $\phi$ of the pair. The signal pair distribution, $S(\Delta\eta,\Delta\phi)$,
represents the yield of particle pairs normalized by $N_\text{trig}$ from the same event,
\begin{equation}
\label{eq:signal}
S(\Delta\eta,\Delta\phi) = \frac{1}{N_\text{trig}}\frac{\rd^{2}N^\text{same}}{\rd\Delta\eta\, \rd\Delta\phi}\ .
\end{equation}
The mixed-event pair distribution,
\begin{equation}
\label{eq:background}
B(\Delta\eta,\Delta\phi) = \frac{1}{N_\text{trig}}\frac{\rd^{2}N^\text{mix}}{\rd\Delta\eta\, \rd\Delta\phi}\ ,
\end{equation}
is constructed by pairing the trigger particles in each event with the
associated particles from 10 different random events in the same 2\unit{cm} wide $z_\text{vtx}$ range
and from the same track multiplicity class.
Here, $N^\text{mix}$ denotes the number of pairs taken from the mixed events.
The ratio $B(0,0)/B(\Delta\eta,\Delta\phi)$ accounts for the
random combinatorial background as well as for pair-acceptance
effects, with $B(0,0)$ representing the mixed-event associated yield for
both particles of the pair going in approximately the same direction and
thus having full pair acceptance (with a bin width of 0.3 in $\Delta\eta$ and
$\pi/16$ in $\Delta\phi$). The signal and background distributions
are first calculated for each event, and then averaged over all the events within
the track multiplicity class. The range of $0<|\deta|<4.8$ and $0<\abs{\dphi}<\pi$ is
used to fill one quadrant of the ($\Delta\eta,\Delta\phi$) histograms, with the
other three quadrants filled (for illustration purposes) by reflection
to cover a ($\Delta\eta,\Delta\phi$) range of $-4.8<\deta<4.8$ and $-\pi/2<\dphi<3\pi/2$ for the 2D
correlation functions, as will be shown later in Fig.~\ref{fig:corr2D_pPb_pt1-3_220260}.

\subsection{Azimuthal Anisotropy Harmonics from Two- and Four-Particle Correlations}

The azimuthal anisotropy harmonics are determined from a Fourier decomposition
of long-range two-particle \dphi\ correlation functions,
\begin{linenomath}
\begin{equation}
\label{eq:Vn}
\frac{1}{N_\text{trig}}\frac{\rd N^\text{pair}}{\rd\Delta\phi} = \frac{N_\text{assoc}}{2\pi} \left[ 1+\sum\limits_{n} 2V_{n\Delta} \cos (n\Delta\phi)\right],
\end{equation}
\end{linenomath}
as described in Refs.~\cite{Chatrchyan:2011eka,Chatrchyan:2012wg}, where $V_{n\Delta}$
are the Fourier coefficients and $N_\text{assoc}$ represents the total number of pairs per trigger
particle for a given $(\pttrg, \ptass)$ bin. The first three Fourier terms are included
in the fits to the dihadron correlation functions. Including additional terms has a
negligible effect on the results of the Fourier fit.
A minimum $|\deta|$ of 2 units is applied to
remove short-range correlations from jet fragmentation. The elliptic and triangular
anisotropy harmonics, $v_{2}\{2,|\deta|>2\}$ and $v_{3}\{2,|\deta|>2\}$,
from the two-particle correlation method can be extracted as a function of \pt\ from the
fitted Fourier coefficients,
\begin{linenomath}
\begin{equation}
\label{eq:Vnpt}
v_{n}\{2,|\deta|>2\}(\pt) = \frac{V_{n\Delta}(\pt,\ptref)}{\sqrt{V_{n\Delta}(\ptref,\ptref)}},~~~~~~~~~~~~~ n=2, 3.
\end{equation}
\end{linenomath}Here, a fixed \ptref\ range for the ``reference particles'' is chosen to be $0.3<\pt<3.0$\GeVc.

The second-order elliptic harmonic, $v_{2}\{4\}$, is also determined from a four-particle cumulant analysis
using the Q-cumulant method described in Ref.~\cite{Bilandzic:2010jr}.
A reference flow $v_{2}\{4\}$ is first determined by correlating four particles
within the tracker acceptance $\abs{\eta}<2.4$ and in a \ptref\ range, $0.3<\ptref<3.0$\GeVc,
\begin{linenomath}
\begin{equation}
\label{eq:v24}
v_{2}^\text{ref}\{4\} = \sqrt[4]{-c_{2}\{4\}} ,
\end{equation}
\end{linenomath}
where the reference four-particle cumulant, $c_{2}\{4\}$, is calculated as,
\begin{linenomath}
\begin{equation}
\label{eq:c24}
c_{2}\{4\}=\left\langle\left\langle \re^{-2i(\phi_{1}+\phi_{2}-\phi_{3}-\phi_{4})}\right\rangle\right\rangle-
2\times\left\langle\left\langle \re^{-2i(\phi_{1}-\phi_{2})}\right\rangle\right\rangle^{2}.
\end{equation}
\end{linenomath}
Here, $\phi_{1}$, $\phi_{2}$, $\phi_{3}$, $\phi_{4}$ are the
azimuthal angles of four different particles in an event,
and $\left\langle\left\langle \cdot \right\rangle\right\rangle$ represents
the average over all particles from all events within a given multiplicity range.

With respect to the reference flow, the differential $v_{2}\{4\}(\pt)$ as a function
of \pt\ is then derived via
\begin{linenomath}
\begin{equation}
\label{eq:v24pt}
v_{2}\{4\}(\pt) = \frac{-d_{2}\{4\}(\pt)}{(v_{2}^\text{ref}\{4\})^{3}},
\end{equation}
\end{linenomath}
where the differential four-particle cumulant, $d_{2}\{4\}(\pt)$, is
calculated by replacing one of the four reference particles in Eq.~(\ref{eq:c24}) by a
particle from a particular \pt\ region. An $\deta$ requirement is not
applied in the four-particle cumulant analysis since short-range two-particle correlations are inherently
minimized by applying this multi-particle method.

\subsection{Corrections and Systematic Uncertainties}
\label{subsec:analysis_systematics}

In performing the correlation analyses, each reconstructed track is
weighted by a correction factor, described in Refs.~\cite{Chatrchyan:2011eka,Chatrchyan:2012wg}.
This factor accounts for the reconstruction efficiency, the detector acceptance, and
the fraction of misreconstructed tracks. Detailed studies of tracking performance
based on MC simulations and collision data can be found in Ref.~\cite{TRK-10-002}.
The combined geometrical acceptance and efficiency for track reconstruction exceeds
60\% for $\pt \approx 0.3$\GeVc and $\abs{\eta}<2.4$. The efficiency is greater than 90\%
in the $\abs{\eta}<1$ region for $\pt>$ 0.6\GeVc. For the entire multiplicity range
(up to \noff\ $\sim$ 350) studied in this paper, no dependence of the
tracking efficiency on multiplicity is found and the rate of misreconstructed tracks
remains at the 1--2\% level.

Based on the studies in Ref.~\cite{TRK-10-002}, the total uncertainty of
the absolute tracking efficiency is estimated to be 3.9\%. This translates directly into a 3.9\%
systematic uncertainty of the extracted associated yields, while the $v_n$ values are
insensitive to it. Systematic uncertainties due to track quality requirements are examined by
varying the track selections for $d_z/\sigma(d_z)$ and $d_{xy}/\sigma(d_{xy})$
from 2 to 5. The results of both associated yields and $v_n$ are found to be insensitive
to these track selections within 2\%. A comparison of high-multiplicity \pPb\ data for a
given multiplicity range but collected by two different HLT triggers with different trigger
efficiencies shows an agreement within 1\%. Possible contamination of residual pileup events
is also investigated. By varying the $z_{vtx}$ range in performing the analysis,
the pileup probability is expected to vary by a factor of 3--4. The systematic uncertainties
for associated yields and $v_n$ from possible residual pileup effects are estimated to be
1--2\% for $\noff\ < 200$, increasing to 6\% for $\noff \geq 260$.

\begin{figure*}[thbp]
  \begin{center}
    \includegraphics[width=\linewidth]{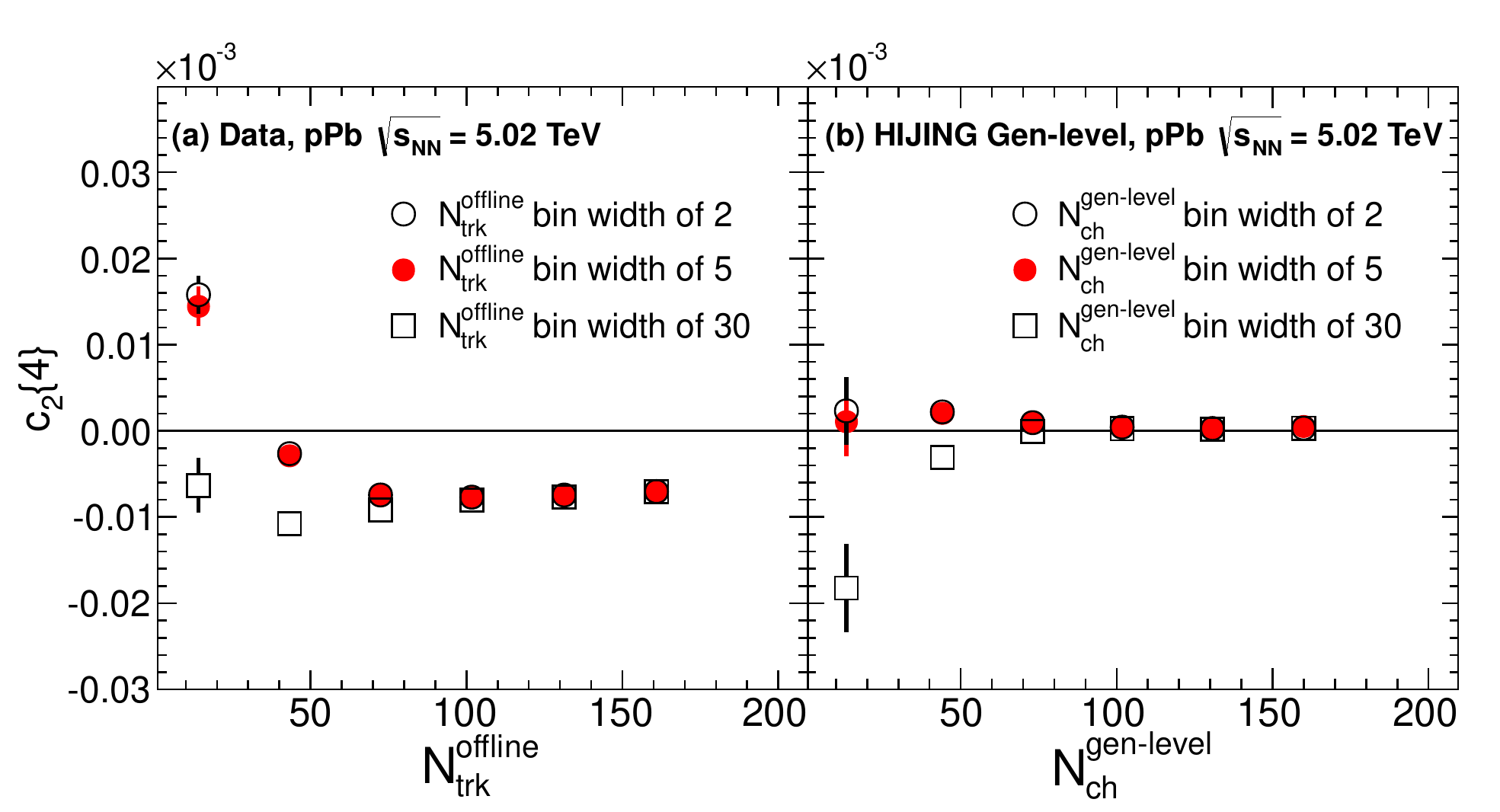}
    \caption{ The $c_{2}\{4\}$ values as a function of multiplicity calculated for
    bin width of 30 (open squares), and first derived using a smaller bin
    width of 2 (open circles) or 5 (solid circles) and then
    averaging over the same wider bin of 30, for \pPb\ \HIJING\ MC simulations (left)
    and data (right) at \rootsNN\ = 5.02\TeV.
    }
    \label{fig:c24}
  \end{center}
\end{figure*}

The event-by-event variation of track multiplicity within a given multiplicity
bin width is found to have an effect on the four-particle cumulant analysis, especially
for the low-multiplicity region. The $c_{2}\{4\}$ values calculated directly for a
multiplicity bin width of 30 show a large discrepancy from those derived first using
a smaller bin width (e.g., 2 or 5) and then averaged over the same wider bin, as
illustrated for \pPb\ data in Fig.~\ref{fig:c24}a and for \pPb\ MC \HIJING\ simulation
(generator-level only) in Fig.~\ref{fig:c24}b. The event multiplicity in \HIJING, $N_\text{ch}^\text{gen-level}$,
is counted for charged primary particles at the generator level with $\abs{\eta}<2.4$ and $\pt > 0.4$\GeVc.
With smaller multiplicity bin widths, the $c_{2}\{4\}$ values
for \HIJING\ are largely consistent with zero. This is expected due to the absence
of collective effect in the \HIJING\ event generator. An \noff\ bin width of 5 is chosen for
the $v_{2}\{4\}$ analysis in this paper. Studies performed with different \noff\ bin widths,
allowing different multiplicity content in the bins, suggest a systematic
uncertainty of only 1\% for $\noff\ > 100$ but up to 10\% for the low multiplicity region $\noff\ < 60$.

The different systematic sources described above are added in quadrature to obtain
the overall systematic uncertainty, shown as boxes in
Figs.~\ref{fig:yieldvspt_new}--\ref{fig:v2_Ntrk_subperiph_atlas}.

\section{Results}
\label{sec:highmult_results}
\subsection{Correlation Functions}
\label{subsec:corrfunc}

\begin{figure}[thb]
\centering
\includegraphics[width=0.45\textwidth]{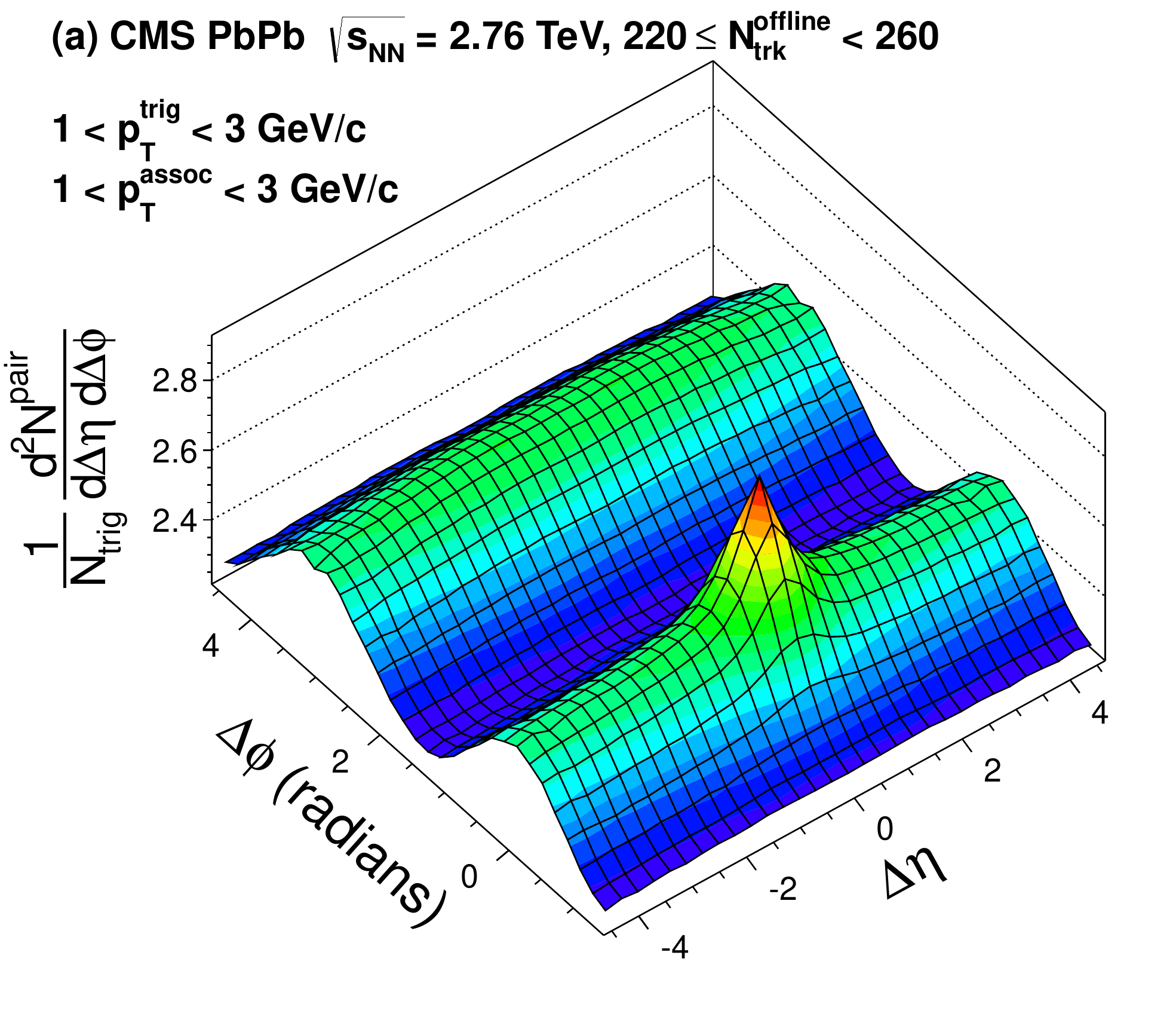}
\includegraphics[width=0.45\textwidth]{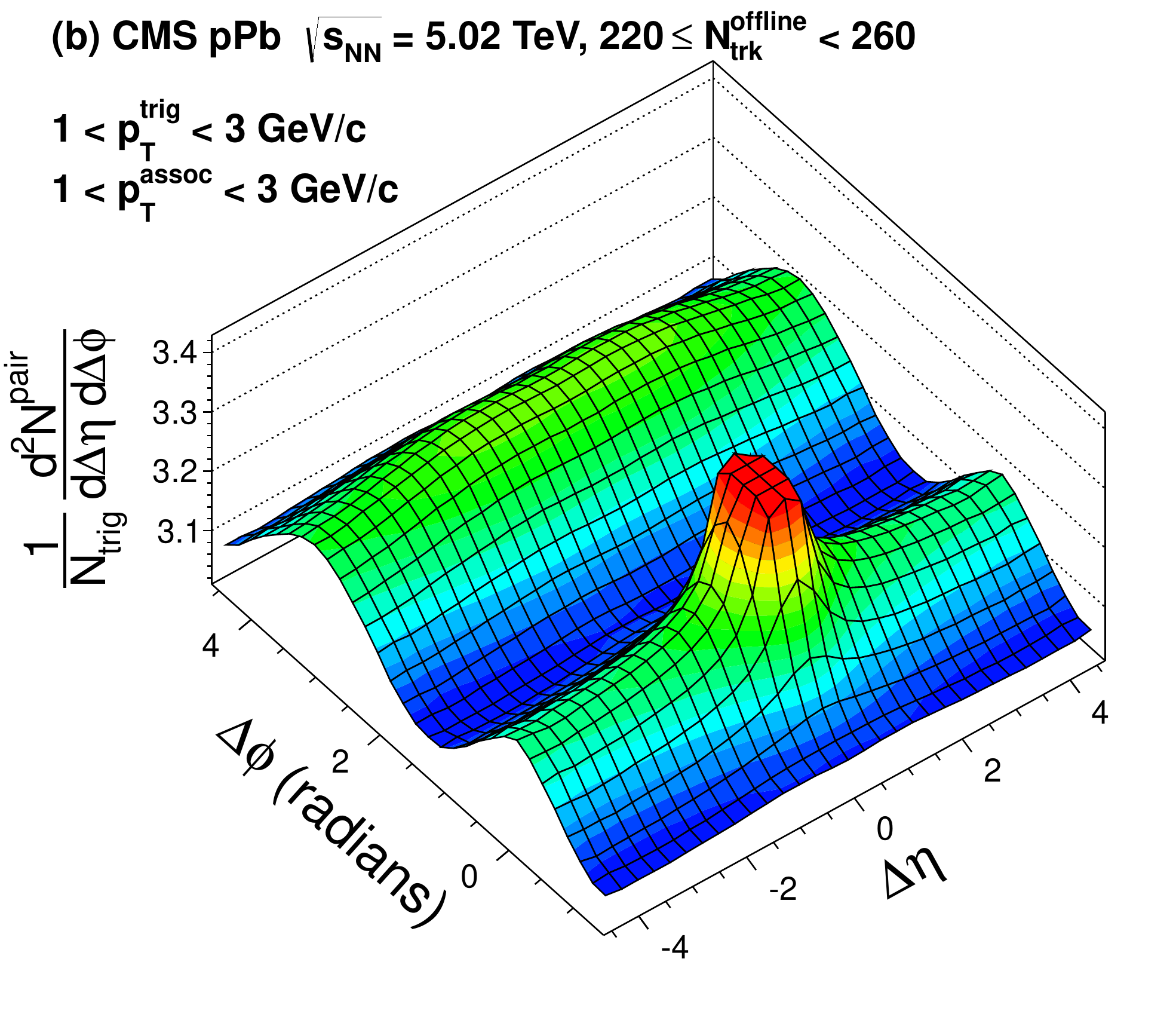}
  \caption{ \label{fig:corr2D_pPb_pt1-3_220260} The 2D two-particle correlation functions
  for (a) 2.76\TeV \PbPb\ and (b) 5.02\TeV \pPb\ collisions for pairs of
  charged particles with $1<\pttrg<3\GeVc$ and $1<\ptass<3\GeVc$ within the $220 \leq \noff\ < 260$
  multiplicity bin. The sharp near-side peak from jet correlations is
  truncated to emphasize the structure outside that region.
   }
\end{figure}

Figure~\ref{fig:corr2D_pPb_pt1-3_220260} shows the 2D two-particle correlation functions
measured in 2.76\TeV \PbPb\ (a) and 5.02\TeV \pPb\ (b) collisions,
for pairs of charged particles with $1<\pttrg<3\GeVc$ and $1<\ptass<3\GeVc$, and with
the track multiplicity in the range $220 \leq \noff< 260$. For \PbPb\ collisions, this \noff\
range corresponds to an average centrality of approximately 60\%,
as shown in Table~\ref{tab:newmultbinning}. For both high-multiplicity systems, in addition to
the correlation peak near $(\deta, \dphi) = (0, 0)$ due to jet fragmentation (truncated
for better illustration of the full correlation structure), a pronounced long-range
structure is seen at $\dphi \approx 0$ extending at least 4.8 units in $|\deta|$.
This structure was previously observed in high-multiplicity ($\noff\ \sim 110$)
\pp\ collisions at \roots\ = 7\TeV~\cite{Khachatryan:2010gv} and \pPb collisions
at \rootsNN\ = 5.02\TeV~\cite{CMS:2012qk,alice:2012qe,atlas:2012fa}.
The structure is also prominent in \AonA collisions over a wide range of energies~\cite{Adams:2005ph,Abelev:2009af,Alver:2008gk,Alver:2009id,Abelev:2009jv,Chatrchyan:2011eka,Chatrchyan:2012wg,Aamodt:2011by,ATLAS:2012at}.
On the away side ($\dphi \approx \pi$) of the correlation functions,
a long-range structure is also seen and found to exhibit a magnitude similar to that
on the near side for this \pt\ range. In non-central \AonA\ collisions, this $\cos(2\Delta\phi)$-like
azimuthal correlation structure is believed to arise primarily from elliptic
flow~\cite{Voloshin:1994mz}. However, the away-side correlations must also contain
contributions from back-to-back jets, which need to be accounted for
before extracting any other source of correlations.

To investigate the observed correlations in finer detail and to obtain a quantitative
comparison of the structure in the \pp, \pPb, and \PbPb\ systems, one-dimensional (1D) distributions in
$\Delta\phi$ are found by averaging the signal and background 2D distributions over
$|\deta| < 1$ (defined as the ``short-range region'') and $|\deta| > 2$ (defined as the ``long-range region'')
respectively, as done in Refs.~\cite{Khachatryan:2010gv,Chatrchyan:2011eka,Chatrchyan:2012wg,CMS:2012qk}.
The correlated portion of the associated yield is estimated using an implementation of the
zero-yield-at-minimum (ZYAM) procedure~\cite{PhysRevC.72.011902}. In this procedure,
the 1D $\Delta\phi$ correlation function is first fitted by a second-order polynomial
in the region $0.1<|\Delta\phi|<2$. The minimum value of the polynomial, $C_\mathrm{ZYAM}$,
is then subtracted from the 1D \dphi\ correlation function
as a constant background (containing no information about correlations) such that its
minimum is shifted to have zero associated yield. The statistical uncertainty
in the minimum level obtained by the ZYAM procedure, combined with the deviations
arising from the choice of fit range in $|\Delta\phi|$, gives an absolute uncertainty
of ${\pm}0.003$ in the associated event-normalized yield that is independent of multiplicity and \pt.

\begin{figure*}[thbp]
\centering
\includegraphics[width=\linewidth]{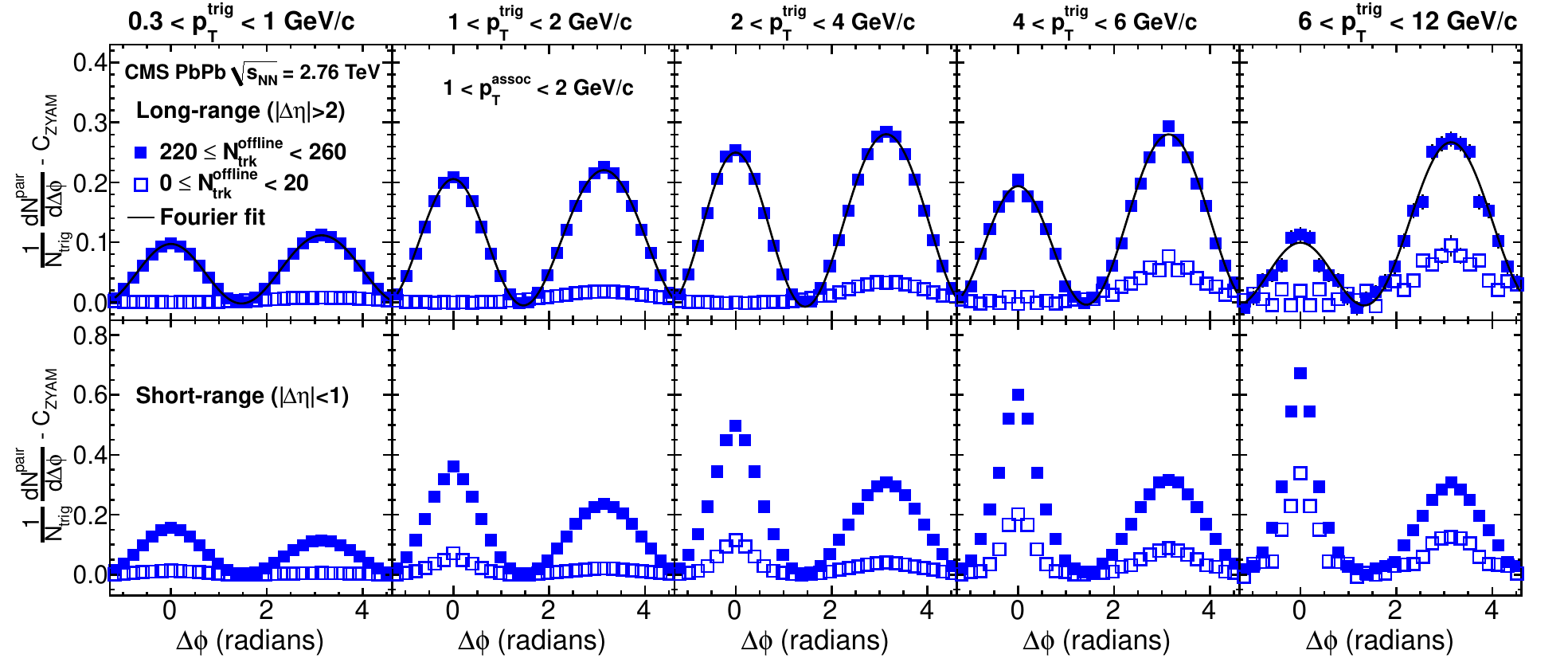}
  \caption{ \label{fig:corr1D_PbPb_N220260_paper} The 1D two-particle correlation
  functions for 2.76\TeV \PbPb\ collisions within the multiplicity range
  $220 \leq \noff\ < 260$ (filled squares) and $\noff\ < 20$ (open squares), for pairs of charged
  particles with fixed \ptass\ 1--2\GeVc in five \pttrg\ ranges,
  in the long-range region ($|\deta|>2$, top) and in the short-range
  region ($|\deta|<1$, bottom). The curves on the top panels correspond to
  the Fourier fits from Eq.~(\ref{eq:Vn}) including the first three terms.
   }
\end{figure*}

\begin{figure*}[thbp]
\centering
\includegraphics[width=\linewidth]{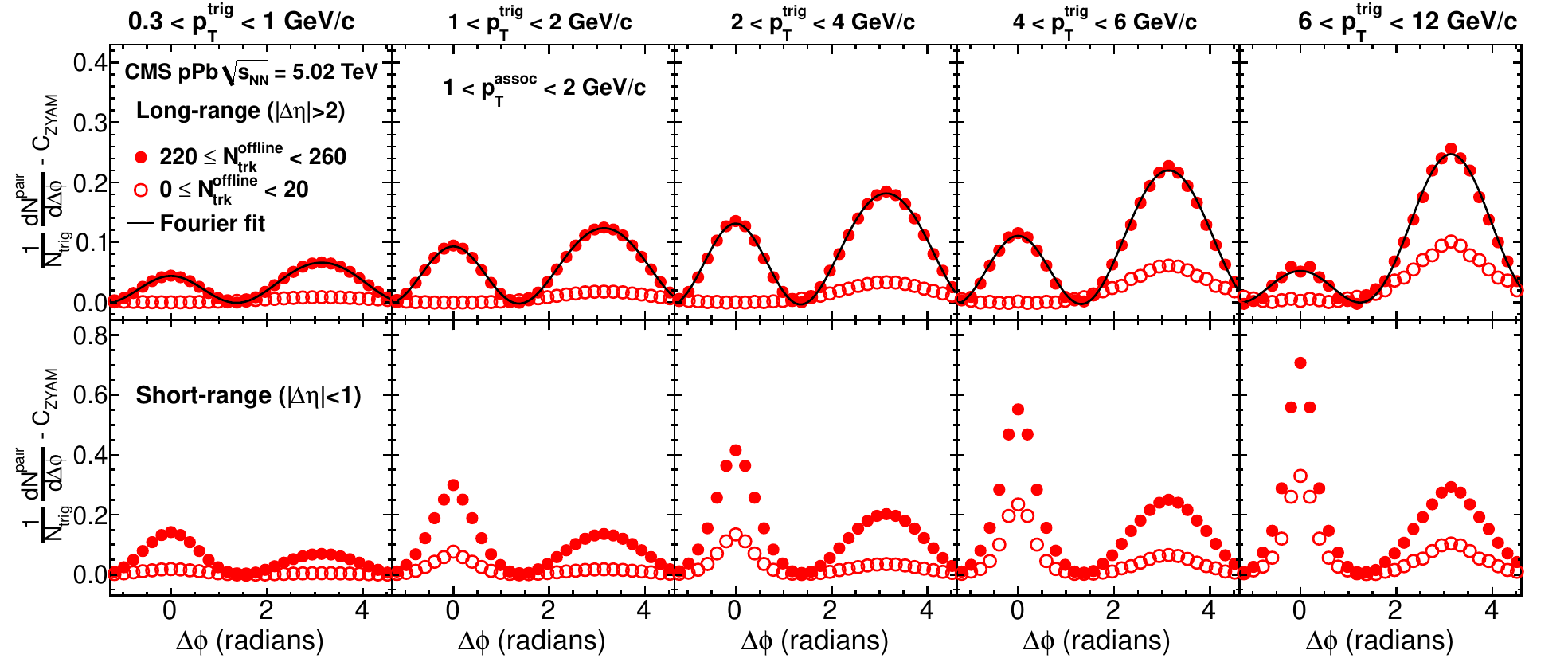}
  \caption{ \label{fig:corr1D_pPb_N220260_paper} The 1D two-particle correlation
  functions for 5.02\TeV \pPb\ collisions under the same conditions
  as in Fig.~\ref{fig:corr1D_PbPb_N220260_paper}.
   }
\end{figure*}

Figures~\ref{fig:corr1D_PbPb_N220260_paper} and \ref{fig:corr1D_pPb_N220260_paper}
show the 1D \dphi\ correlation functions, after applying the ZYAM procedure,
for \PbPb\ and \pPb\ data, respectively, in the multiplicity range
$\noff\ < 20$ (open) and $220 \leq \noff\ < 260$ (filled). Various selections of \pttrg\
are shown for a fixed \ptass\ range of 1--2\GeVc in both the long-range (top) and
short-range (bottom) regions, with \pt\ increasing from left to right. The curves
show the Fourier fits from Eq.~(\ref{eq:Vn}), which will be discussed
in detail later. The \pPb\ and \PbPb\ yields show a similar correlation structure and a similar
evolution of this structure with \pttrg\ over a wide range of \pttrg.
As illustrated in Fig.~\ref{fig:corr2D_pPb_pt1-3_220260}, while
the near-side long-range signal varies only by a small amount over almost 5 units in
$\Delta\eta$, the short-range region shows a strong $\Delta\eta$ dependence. Therefore,
the \dphi\ correlation functions in the short-range region of Figs.~\ref{fig:corr1D_PbPb_N220260_paper}
and \ref{fig:corr1D_pPb_N220260_paper} reflect the contributions of both jet fragmentation
and long-range correlations. For $\noff\ < 20$, no near-side correlations are
observed in the long-range region of either \pPb\ or \PbPb\ data.

\subsection{Integrated Associated Yields}
\label{subsec:yield}

\begin{figure*}[thbp]
\centering
\includegraphics[width=\textwidth]{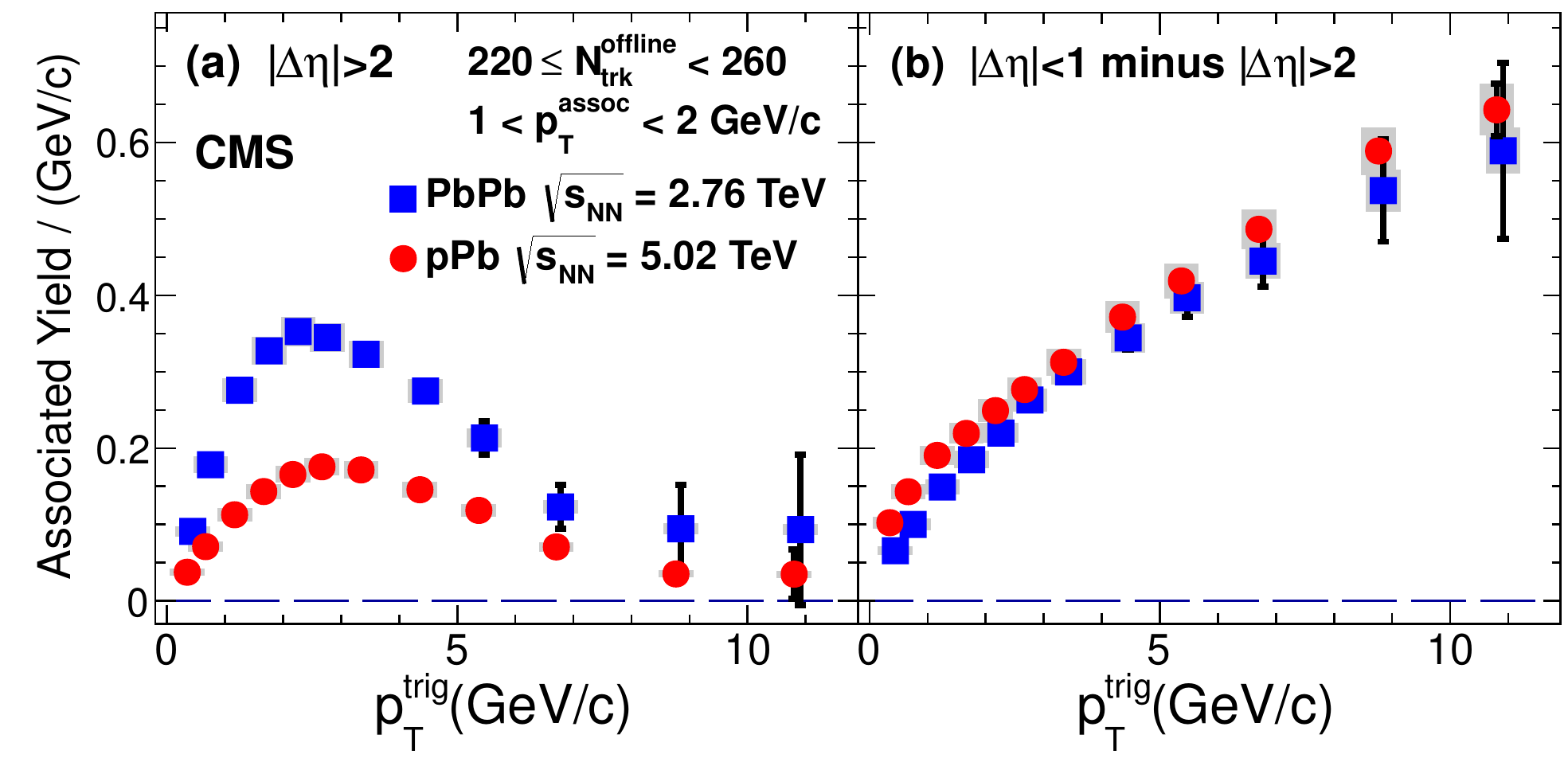}
  \caption{ \label{fig:yieldvspt_new} Associated event-normalized yield for the near-side
  correlation function integrated over the region $\abs{\dphi}<1.2$,
  averaged over the (a) long-range ($|\deta|>2$) region and (b) short-range ($|\deta|<1$) region,
  from which the event-normalized yield of the long-range region is subtracted. The results are shown
  as a function of \pttrg\ at $1<\ptass<2$\GeVc for events with $220 \leq \noff\ < 260$
  for 5.02\TeV \pPb\ collisions (filled circles) and 2.76\TeV \PbPb\ collisions (filled squares).
  The error bars correspond to statistical uncertainties, while the shaded areas denote
  the systematic uncertainties.
   }
\end{figure*}

\begin{figure*}[thbp]
\centering
\includegraphics[width=\textwidth]{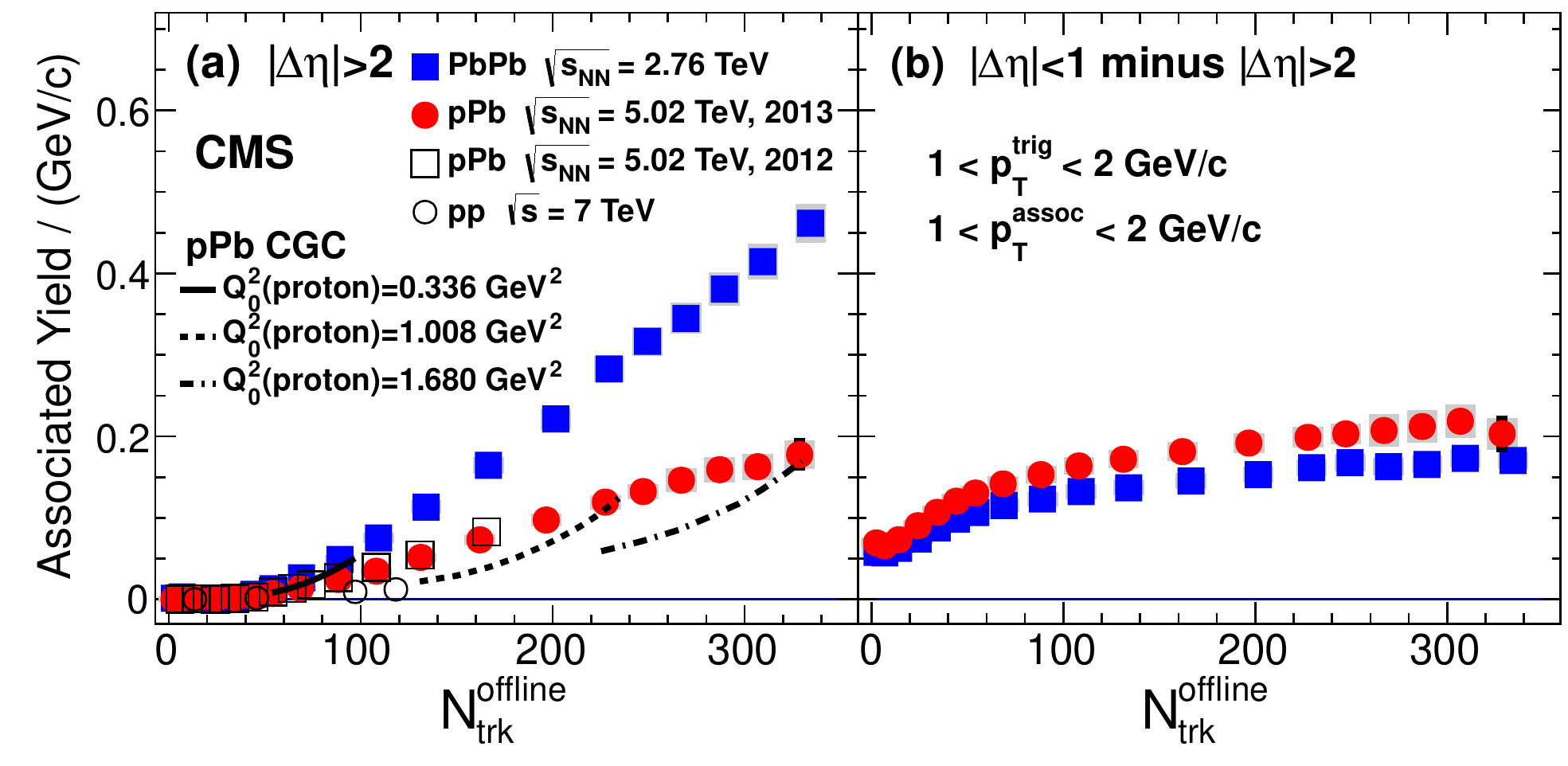}
  \caption{ \label{fig:yieldvsmult_new} Associated event-normalized yields for the near-side
  correlation function as a function of multiplicity \noff for $1<\pttrg<2\GeVc$ and $1<\ptass<2\GeVc$
  under the same conditions as in Fig.~\ref{fig:yieldvspt_new}.
  The results for 7\TeV \pp\ collisions (open circles)~\cite{Khachatryan:2010gv}
  and 5.02\TeV \pPb\ collisions from 2012 run (open squares)~\cite{Chatrchyan:2012wg},
  as well as calculations from the Color Glass Condensate (CGC) theory (curves)~\cite{Dusling:2013}, are also shown.
   }
\end{figure*}

The strength of the near-side correlations for short- and long-range regions
can be further quantified by integrating the event-normalized associated yield from
Figs.~\ref{fig:corr1D_PbPb_N220260_paper} and \ref{fig:corr1D_pPb_N220260_paper}
over $\abs{\dphi} < 1.2$. The resulting integrated yields are shown for \pPb\ and \PbPb\
in Fig.~\ref{fig:yieldvspt_new} as a function of \pttrg\ for $1<\ptass<2$\GeVc and $220\leq \noff < 260$,
and in Fig.~\ref{fig:yieldvsmult_new} as a function of \noff\ for $1<\pttrg<2$\GeVc and $1<\ptass<2$\GeVc
together with the \pp\ results from Ref.~\cite{Khachatryan:2010gv}.
The ``jet yield'' is extracted by subtracting the event-normalized integrated yield in the long-range region
from that in the short-range region. The error bars correspond to statistical
uncertainties, while the shaded boxes indicate the systematic uncertainties discussed
in Section~\ref{subsec:analysis_systematics}.

The jet yield (Fig.~\ref{fig:yieldvspt_new}b) increases with \pttrg\ in both \pPb\
and \PbPb\ as would be expected if higher-energy jets, which fragment
into more final-state particles, are selected by requiring higher-\pttrg\ particles.
In striking contrast to the jet yields, the \pttrg\ dependence of the
long-range yields (Fig.~\ref{fig:yieldvspt_new}a)
show an initial rise, reaching a maximum at $\pt \approx 2$--$3\GeVc$,
followed by a falloff with values consistent with zero for \pttrg\ $\sim$ 12\GeVc.

The jet yield (shown in Fig.~\ref{fig:yieldvsmult_new}b) as a function of multiplicity
increases by a factor of two as \noff\ increases from 0 to 60. It
then rises moderately by 20--30\%, for $60 \leq \noff\ < 350$,
the limit of this measurement. This demonstrates that by selecting high-track-multiplicity
\PbPb\ and \pPb\ events, there is no significant bias to stronger jet-like correlations
(at least for the \pt\ range of 1--2\GeVc). It was previously observed in Ref.~\cite{CMS:2012qk}
that the long-range yield as a function of multiplicity only becomes significant at $\noff \sim 40$--$50$,
followed by a monotonic rise with \noff\ in \pp\ and \pPb\ collisions.
In this paper, the measurement of the long-range yield (Fig.~\ref{fig:yieldvsmult_new}a) in \pPb\ collisions
is extended to a significantly wider multiplicity range. A direct comparison to the \pp\ \cite{Khachatryan:2010gv}
and \PbPb\ collision systems is also provided. The \PbPb\ long-range yield is found to become significant
for $\noff\ \gtrsim 40$--$50$, similar to the \pp\ and \pPb\ results. For both \pPb\ and \PbPb\ data, the
long-range yields continue increasing with multiplicity up to $\noff \sim 350$. The long-range yield
in \PbPb\ is about a factor of two larger than in \pPb, and a factor of eight larger
than in \pp\ at a given multiplicity and \pttrg\ value.
In contrast to the weak multiplicity dependence of jet-like correlations shown in
Fig.~\ref{fig:yieldvsmult_new}b at higher values of \noff, a monotonic increase
of the magnitude of the long-range yield with the overall event multiplicity is observed
in all three collision systems.

In the framework of the color-glass condensate model, the long-range correlation structure
in \pPb\ collisions has been attributed to initial-state gluon correlations,
where the contribution of collimated gluon emissions is significantly enhanced in the gluon
saturation regime~\cite{Dusling:2012cg,Dusling:2012wy,Dusling:2013}. This model qualitatively
describes the increase in the long-range yield for higher-multiplicity events
as shown in Fig.~\ref{fig:yieldvsmult_new}a, where three different initial proton
saturation scales are assumed for the \pPb system. Since the calculations depend
on saturation scales for both protons and lead nuclei, the data provide valuable
constraints on the multiplicity dependence of these parameters in the model.

\subsection{Fourier Harmonics \texorpdfstring{$v_n$}{v[n]}}
\label{subsec:vn}

\begin{figure*}[thb]
  \begin{center}
    \includegraphics[width=\textwidth]{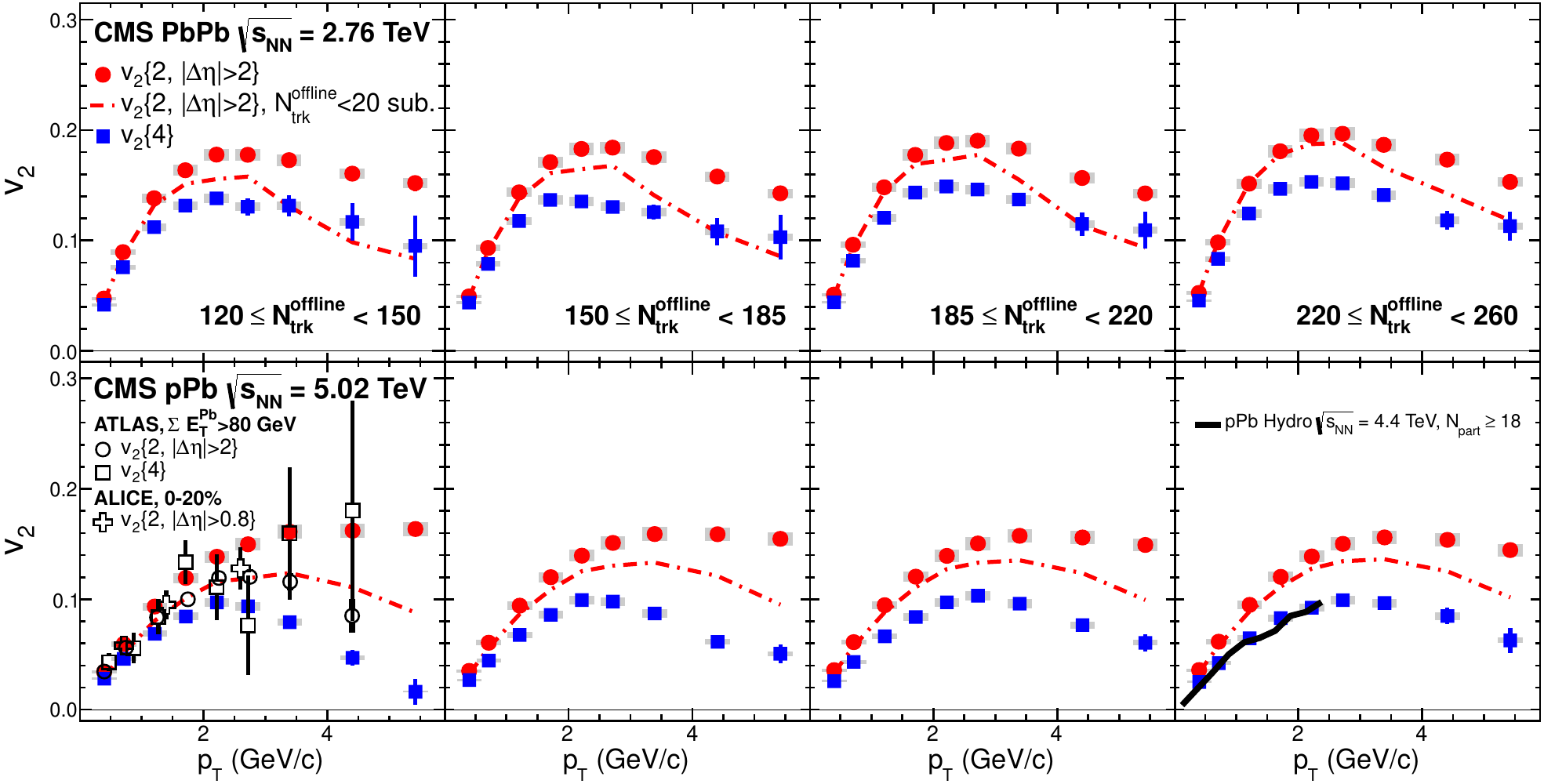}
    \caption{
    The differential $v_2\{2, \abs{\Delta\eta} > 2\}$ (filled circles)
    and $v_{2}\{4\}$ (filled squares) values for four multiplicity ranges
    obtained with $\abs{\eta}<2.4$ and a \ptref\ range of 0.3--3\GeVc. The results are for
    2.76\TeV \PbPb\ collisions (top) and for 5.02\TeV \pPb\ collisions (bottom).
    The error bars correspond to statistical uncertainties, while the shaded areas denote
    the systematic uncertainties. Results after subtracting the low-multiplicity data ($\noff<20$)
    as well as predictions from a hydrodynamic model are also shown (curves). The open
    markers show the results from ALICE~\cite{alice:2012qe} and ATLAS~\cite{Aad:2013fja}
    using 2012 \pPb\ data.
    }
    \label{fig:v2_pt8panel}
  \end{center}
\end{figure*}

\begin{figure*}[thb]
  \begin{center}
    \includegraphics[width=\textwidth]{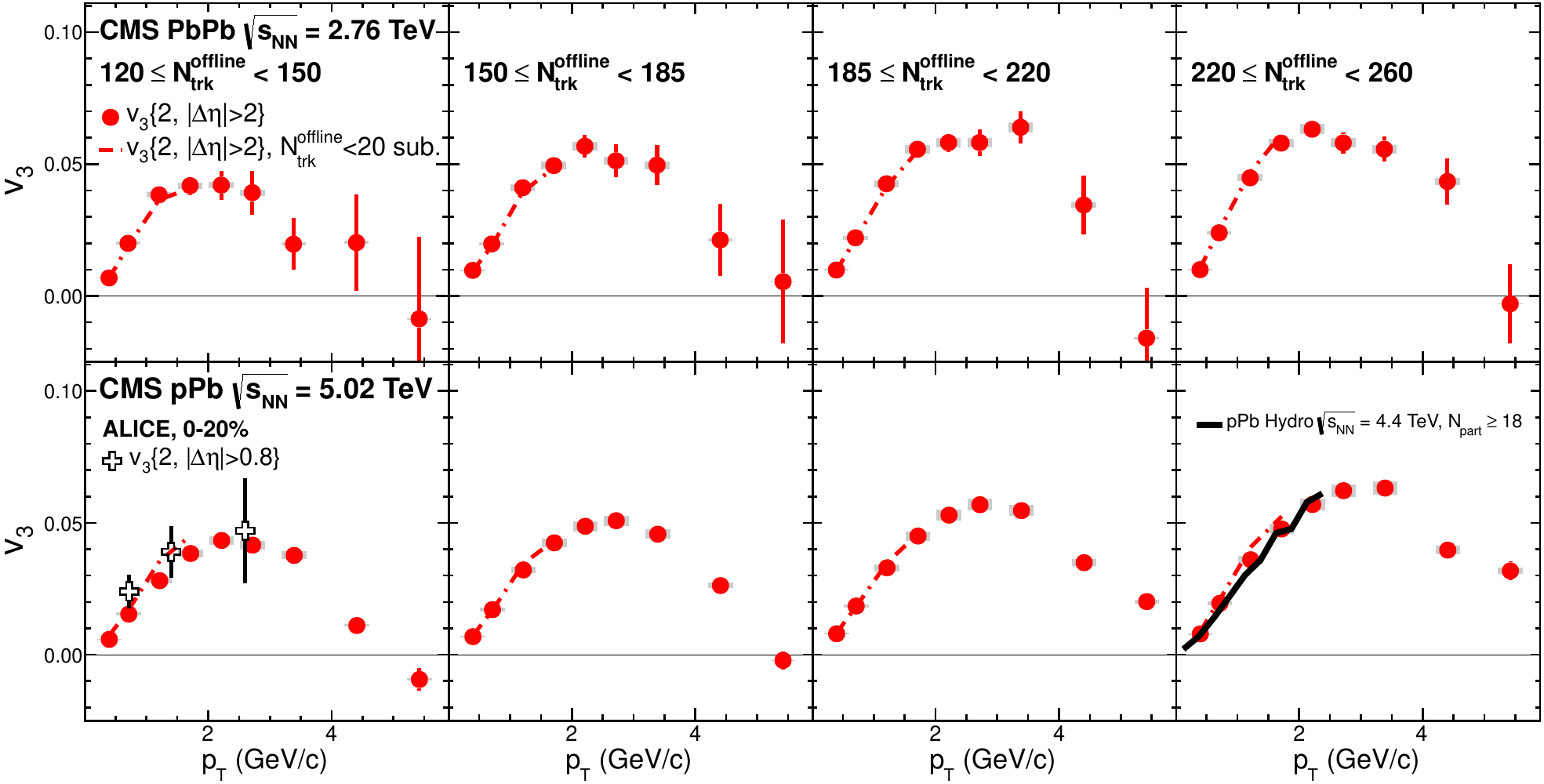}
    \caption{
    The differential $v_3\{2, \abs{\Delta\eta} > 2\}$ values for four multiplicity ranges
    under the same conditions as in Fig.~\ref{fig:v2_pt8panel}.
    }
    \label{fig:v3_pt8panel}
  \end{center}
\end{figure*}

Long-range correlations in \pPb\ collisions have also been predicted in hydrodynamic
models~\cite{Bozek:2011if} where a collective hydrodynamic expansion of the system with fluctuating
initial conditions is assumed. To compare with hydrodynamic predictions of the long-range
correlations in \pPb\ collisions, the elliptic ($v_2$) and
triangular ($v_3$) flow harmonics are extracted from a Fourier decomposition of 1D $\Delta\phi$
correlation functions, $v_2\{2, \abs{\Delta\eta} > 2\}$ and $v_3\{2, \abs{\Delta\eta} > 2\}$,
for the long-range region ($\abs{\Delta\eta} > 2$) as shown in Figs.~\ref{fig:v2_pt8panel}
and \ref{fig:v3_pt8panel}, respectively. To further reduce the
residual nonflow correlations on the away side, a four-particle cumulant analysis is
also used to extract the elliptic flow, $v_2\{4\}$, as shown in Fig.~\ref{fig:v2_pt8panel}.
As mentioned in Section~\ref{sec:intro}, the multi-particle correlation technique has the
advantage of suppressing short-range jet-like correlations compared to two-particle
correlations. It thus provides a cleaner measurement of the long-range correlations
of collective nature involving many particles from the system.

As seen in Fig.~\ref{fig:v2_pt8panel}, the magnitude of the $v_2$
signal is found to be larger in \PbPb\ than in \pPb by about 30\% for $\pt<2$\GeVc
(the near-side long-range yield is related to $v_2^{2}$ as suggested in Eq.~(\ref{eq:Vn}), and
thus differs by a larger factor between the two systems as shown in Fig.~\ref{fig:yieldvsmult_new}).
The difference between the $v_{2}\{2, |\deta|>2\}$ and $v_2\{4\}$ results
could be, a consequence of event-by-event fluctuations in the flow signal or nonflow correlations,
as believed to be the case in \PbPb\ collisions~\cite{Ollitrault:2009ie}.
The $v_{3}\{2, |\deta|>2\}$ component, shown in Fig.~\ref{fig:v3_pt8panel}, reaches the same
maximum value for the two systems but has a much smaller magnitude than $v_{2}\{2, |\deta|>2\}$
over the entire \pt\ range investigated here. The \pt\ dependencies of both the $v_2$ and $v_3$ coefficients
are similar, with peak values at 2--3\GeVc range for \PbPb\ and slightly higher
for \pPb. The elliptic- and triangular-flow components predicted by the hydrodynamic
calculation of Ref.~\cite{Bozek:2011if} for \pPb\ collisions at \rootsNN\ = 4.4\TeV and
for $\pt<2.5$\GeVc are also shown, and compared to the high-multiplicity \pPb\ data in
Figs.~\ref{fig:v2_pt8panel} and \ref{fig:v3_pt8panel}. The calculations have little
collision energy dependence, and assume the number of participating nucleons to be larger
or equal to 18, approximately corresponding to the top 4\% central \pPb\ events. However,
contributions from event-by-event fluctuations of the flow signal around its average value
are not accounted for in the calculations. Therefore, the $v_2$ calculated in Ref.~\cite{Bozek:2011if}
is expected to lie between the values from the two- and four-particle correlation
methods~\cite{Ollitrault:2009ie}. Detailed studies of $v_2$ using various techniques
in \PbPb\ collisions at \rootsNN\ = 2.76\TeV by CMS can be found in Ref.~\cite{Chatrchyan:2012ta}.

As mentioned above, the residual jet-like correlations on the away side of the two-particle
correlation function could contribute to the extracted $v_{n}\{2, |\deta|>2\}$ signal,
and thus induce a systematic uncertainty in the quantitative comparison to hydrodynamic calculations.
Assuming that the jet-induced correlations are invariant with event multiplicity
in \pPb\ collisions, the ALICE~\cite{alice:2012qe} and ATLAS~\cite{atlas:2012fa} experiments
proposed to subtract the results of low-multiplicity events,
where the long-range correlation signal is not present, from those of high-multiplicity events.
While further justification of this assumption is still required, a similar procedure
is applied in this paper for comparison purposes. The Fourier coefficients, $V_{n\Delta}$,
extracted from Eq.~(\ref{eq:Vn}) for $\noff\ < 20$ (corresponding to the 70--100\%
lowest-multiplicity events for \pPb) are subtracted from the data in the higher-multiplicity region:
\ifthenelse{\boolean{cms@external}}{\begin{linenomath}
\begin{multline}
\label{eq:vnsubperiph}
V^\text{ sub}_{n\Delta}=V_{n\Delta}-V_{n\Delta}(\noff<20)\times\\\frac{N_\text{assoc}(\noff<20)}{N_\text{assoc}}\times
\frac{Y_\text{jet}}{Y_\text{jet}(\noff<20)},
\end{multline}
\end{linenomath}
}
{
\begin{linenomath}
\begin{equation}
\label{eq:vnsubperiph}
V^\text{sub}_{n\Delta}=V_{n\Delta}-V_{n\Delta}(\noff<20)\times\frac{N_\text{assoc}(\noff<20)}{N_\text{assoc}}\times\frac{Y_\text{jet}}{Y_\text{jet}(\noff<20)},
\end{equation}
\end{linenomath}
}
where $Y_\text{jet}$ represents the near-side jet yield.
The ratio, $Y_\text{jet}/Y_\text{jet}(\noff<20)$, is introduced to account for
the enhanced jet correlations due to the selection of higher-multiplicity events
seen in Fig.~\ref{fig:yieldvsmult_new}b. This procedure is tested using the \HIJING\ model,
where there are no final-state interactions of jets in \pPb\ collisions.
The residual $V^\text{sub}_{n\Delta}$ in \HIJING\ after subtraction
is found to be less than 5\%. The low-multiplicity-subtracted $v_{2}\{2, |\deta|>2\}$ and
$v_{3}\{2, |\deta|>2\}$ (limited to $\pt<2$\GeVc for $v_3$ due to the low statistical precision of the
low-multiplicity data) are shown as dash-dotted curves in Figs.~\ref{fig:v2_pt8panel} and \ref{fig:v3_pt8panel}.
After applying the subtraction procedure, the results at low \pt\ remain almost unchanged,
while a reduction is seen in $v_2$ for higher \pt\ particles. This is consistent with the
observation of stronger jet-like correlations at higher \pt\ in Fig.~\ref{fig:yieldvspt_new}b.
The CMS data are compared to the measurement by the ATLAS experiment for an event
multiplicity class (selected based on the total transverse energy measured with $3.1<\eta<4.9$
in the direction of the \Pb\ beam) comparable to $120 \leq \noff < 150$ used in the CMS analysis,
after subtracting the 50--100\% lowest-multiplicity data. The $v_2\{2\}$ and $v_3\{2\}$ data
measured by the ALICE experiment for the 0--20\% highest-multiplicity \pPb\ collisions~\cite{alice:2012qe}
are also shown in Figs.~\ref{fig:v2_pt8panel} and \ref{fig:v3_pt8panel}. Results from all
three experiments are consistent within quoted uncertainties.

\begin{figure*}[thb]
  \begin{center}
    \includegraphics[width=\textwidth]{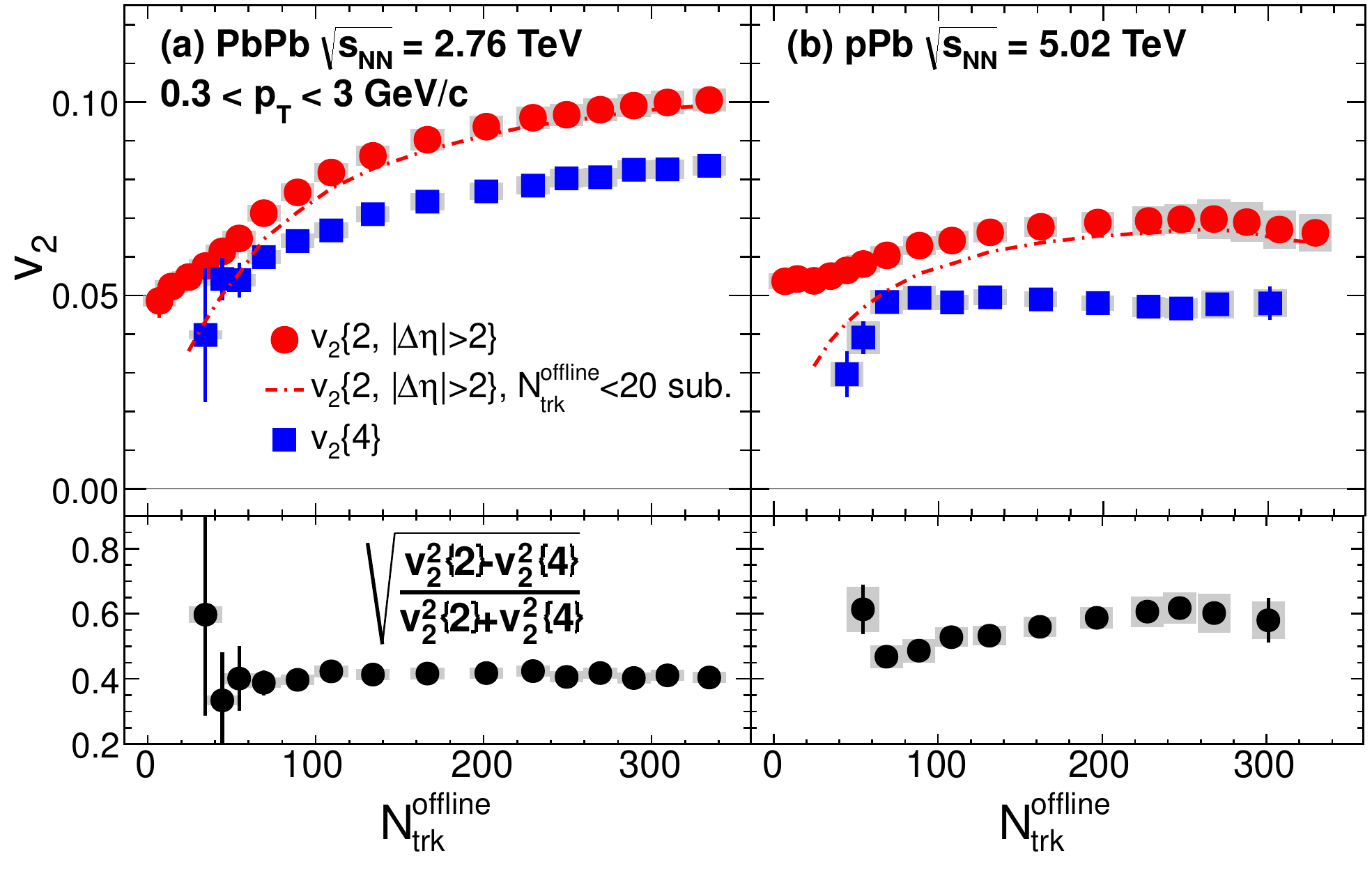}
    \caption{ Top: the $v_2\{2, \abs{\Delta\eta} > 2\}$ (circles) and $v_{2}\{4\}$ (squares) values as a
    function of \noff for $0.3<\pt<3$\GeVc, in 2.76\TeV \PbPb\ collisions (left) and 5.02\TeV \pPb\
    collisions (right). Bottom: upper limits on the relative $v_2$ fluctuations
    estimated from $v_{2}\{2\}$ and $v_{2}\{4\}$ in 2.76\TeV \PbPb\ collisions (left) and
    5.02\TeV \pPb\ collisions (right). The error bars correspond to statistical uncertainties,
    while the shaded areas denote the systematic uncertainties. Results after subtracting
    the low-multiplicity data ($\noff<20$) are also shown (curves).
    }
    \label{fig:v2_Ntrk_subperiph}
  \end{center}
\end{figure*}

\begin{figure*}[thb]
  \begin{center}
    \includegraphics[width=\textwidth]{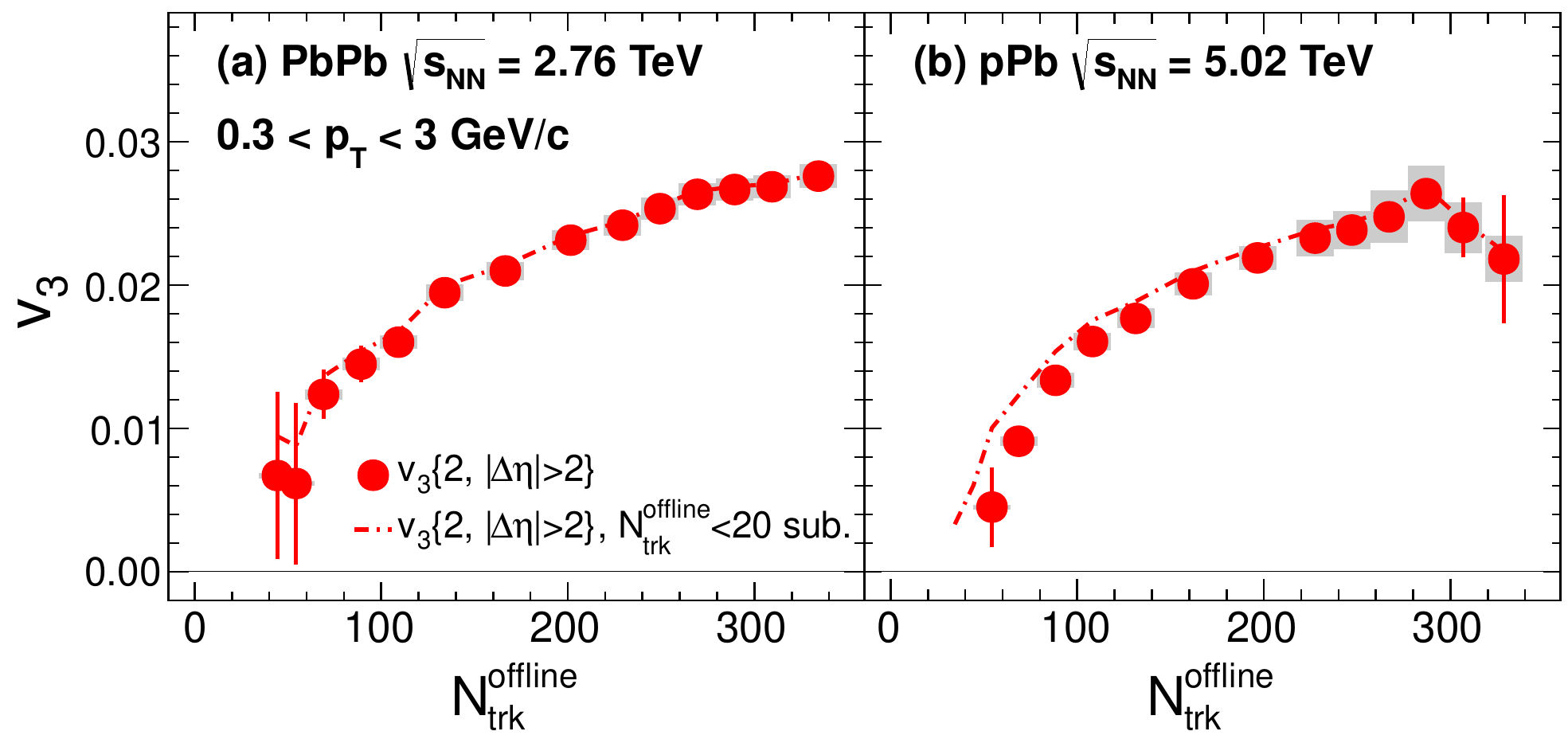}
    \caption{
     The $v_3\{2, \abs{\Delta\eta} > 2\}$ values as a function of \noff for $0.3<\pt<3$\GeVc,
     in 2.76\TeV \PbPb\ collisions (left) and 5.02\TeV \pPb\ collisions (right).
     The error bars correspond to statistical uncertainties, while the shaded areas
     denote the systematic uncertainties.
    }
    \label{fig:v3_Ntrk_subperiph}
  \end{center}
\end{figure*}

\begin{figure}[thb]
  \begin{center}
    \includegraphics[width=\cmsFigWidth]{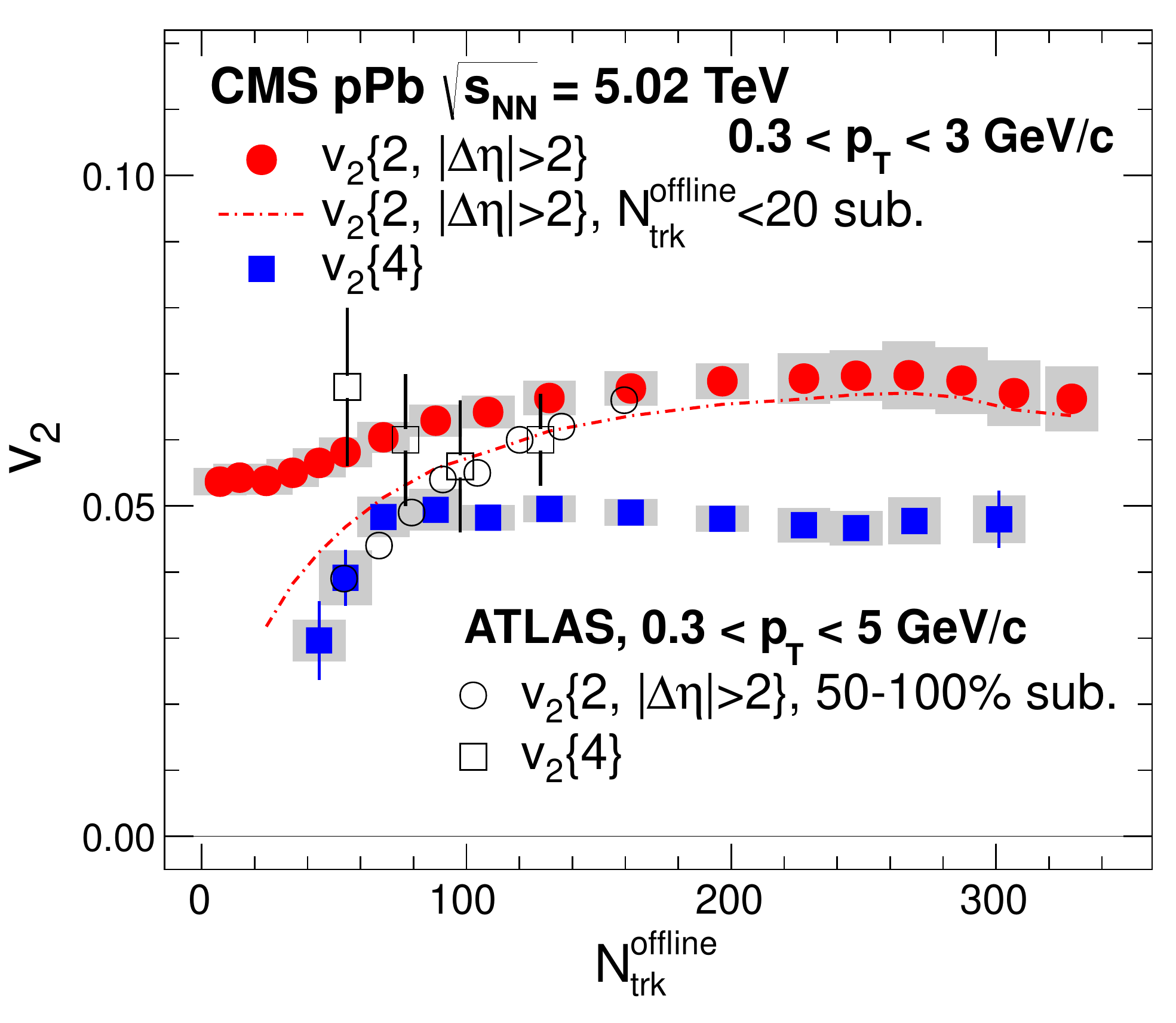}
    \caption{ The $v_2\{2, \abs{\Delta\eta} > 2\}$ and $v_{2}\{4\}$ values as a
    function of \noff for $0.3<\pt<3$\GeVc, measured by CMS in 5.02\TeV \pPb\
    collisions (filled). The dash-dotted curve shows the CMS $v_2\{2, \abs{\Delta\eta} > 2\}$
    values after subtracting the 70--100\% lowest-multiplicity data, to be
    compared with the ATLAS results subtracted by 50-100\% lowest-multiplicity
    data (open)~\cite{Aad:2013fja}. The error bars correspond to statistical uncertainties,
    while the shaded areas denote the systematic uncertainties.
    }
    \label{fig:v2_Ntrk_subperiph_atlas}
  \end{center}
\end{figure}

The multiplicity dependencies of $v_2$ and $v_3$ for \PbPb\ and \pPb\ collisions,
averaged over the \pt\ range from 0.3 to 3\GeVc, are shown in
Figs.~\ref{fig:v2_Ntrk_subperiph} and \ref{fig:v3_Ntrk_subperiph}, respectively.
The $v_{2}\{2, |\deta|>2\}$ and $v_{2}\{4\}$ values in \PbPb\ collisions exhibit a
moderate increase with \noff, while these coefficients remain relatively constant
as a function of multiplicity for \pPb\ data at larger values of \noff.
This is consistent with the monotonic rise of the associated yield as a
function of multiplicity shown in Fig.~\ref{fig:yieldvsmult_new}, which is
mainly driven by the increase of total number of pairs per trigger particle,
as indicated in Eq.~(\ref{eq:Vn}).
Similarly to Figs.~\ref{fig:v2_pt8panel} and \ref{fig:v3_pt8panel}, the \PbPb\ data
show a larger $v_2$ signal than observed for the \pPb\ data over a wide multiplicity
range, while the magnitude of $v_{3}\{2, |\deta|>2\}$ is remarkably similar for both
systems at the same event multiplicity.
This similarity of the triangular flow is not trivially expected within a
hydrodynamic picture since the initial-state collision geometry
is very different for the \pPb\ and \PbPb\ systems.
Below an \noff\ value of 40--50, neither $v_{3}\{2, |\deta|>2\}$
nor $v_{2}\{4\}$ could be reliably extracted. The loss of a $v_{2}\{4\}$ signal
indicates either the absence of collective effects for very-low-multiplicity
collisions, or the breakdown of the four-particle cumulant technique
in the limit of a small number of particles. The procedure of subtracting the
low-multiplicity data to attempt to remove jet correlations is also performed here and
shown as dash-dotted curves in Figs.~\ref{fig:v2_Ntrk_subperiph} and \ref{fig:v3_Ntrk_subperiph}.
The $v_{3}\{2, |\deta|>2\}$ values become larger after subtraction, especially for
the low-multiplicity region, due to the fact that $V_{3\Delta}$
extracted for $\noff<20$ is negative.
The resulting $v_{2}\{2, |\deta|>2\}$ and $v_{3}\{2, |\deta|>2\}$ are
found to remain almost unchanged after subtraction in the high-multiplicity region (i.e., for $\noff>200$).
This is expected since, for a given associated yield from jet correlations, the
contribution to $v_{n}\{2\}$ is suppressed by $1/\sqrt{\smash[b]{\noff}}$ as the multiplicity
increases, as indicated by Eq.~(\ref{eq:Vn}). Therefore, the higher-multiplicity
events provide a much cleaner environment for studying the long-range correlations.

Figure~\ref{fig:v2_Ntrk_subperiph_atlas} shows the comparison of $v_{2}\{2, |\deta|>2\}$
and $v_{2}\{4\}$ results as a function of multiplicity from CMS, averaged over $0.3<\pt<3$\GeVc,
with those obtained by the ATLAS experiment, averaged over $0.3<\pt<5$\GeVc with the data
from the 2012 \pPb\ run. The ATLAS $v_{2}\{2, |\deta|>2\}$
values have the contribution from the 50--100\% lowest multiplicity data subtracted,
while the corresponding CMS data, shown as a curve in Fig.~\ref{fig:v2_Ntrk_subperiph_atlas},
use the 70--100\% lowest multiplicity events for the subtraction. The difference in the
low-multiplicity events used for the subtraction could explain the slight
discrepancy in the resulting $v_{2}\{2, |\deta|>2\}$ data from the two experiments.
The $v_{2}\{4\}$ values from ATLAS
are systematically higher than the CMS data. This may be accounted for by the multiplicity
fluctuation effect discussed previously (e.g., Fig.~\ref{fig:c24}), although the discrepancy
is not large with respect to the uncertainties.

Finally, the magnitude of event-by-event $v_2$
fluctuations is estimated from the difference in the $v_{2}\{2, |\deta|>2\}$
and $v_{2}\{4\}$ results. If hydrodynamic flow is the dominant source of the correlations,
the relative $v_2$ fluctuations can be approximated by $\sqrt{\smash[b]{(v_{2}^{2}\{2\}-v_{2}^{2}\{4\})/(v_{2}^{2}\{2\}+v_{2}^{2}\{4\})}}$~\cite{Ollitrault:2009ie}.
The resulting flow fluctuation values calculated for \pPb\ and \PbPb\ collisions are
shown in the bottom two panels of Fig.~\ref{fig:v2_Ntrk_subperiph}, with
40\% $v_2$ fluctuations observed in \PbPb\ and 50--60\% fluctuations in \pPb\ collisions.
This magnitude of $v_2$ fluctuations in 2.76\TeV \PbPb\ collisions at the LHC is
comparable to the value measured in 200\GeV \AuAu\ collisions at RHIC~\cite{Alver:2007qw}.
As a consequence of possible residual nonflow correlations
from back-to-back jets on the away side in the $v_{2}\{2, |\deta|>2\}$
measurement, these results should be considered as upper limits on the
flow fluctuations.

\section{Summary}
\label{sec:conclusion}

Detailed studies of two- and four-particle azimuthal correlations have been
performed in
\pPb\ collisions at \rootsNN\ = 5.02\TeV by the CMS experiment. The new measurements
extend previous CMS two-particle correlation analyses in \pPb\ collisions
to a significantly broader particle multiplicity range.
A direct comparison of the correlation data between \pPb\ and \PbPb\ collisions
was presented as a function of particle multiplicity and transverse momentum.
The observed correlations were quantified in terms of the integrated near-side
associated yields and azimuthal anisotropy Fourier harmonics ($v_n$).
For both \pPb and \PbPb\ collisions, elliptic ($v_2$) and triangular ($v_3$)
flow Fourier harmonics were extracted from long-range two-particle correlations.
Furthermore, the elliptic flow was studied with a four-particle cumulant analysis,
where multi-particle correlations can be directly investigated.

For a fixed \ptass\ range, the long-range yield and anisotropy harmonics
show similar trends as a function of \pttrg, first increasing and then decreasing
with a maximum at \pttrg\ $\approx$2--3\GeVc in both \pPb\ and \PbPb\ collisions.
For \pPb\ collisions, the long-range associated yield rises monotonically with
particle multiplicity. Correspondingly, the $v_2$ harmonics obtained from
the two- and four-particle correlation analyses show only a weak multiplicity dependence.
Comparing the \pPb\ and \PbPb\ systems at the same
multiplicity and \pt, the long-range yield and $v_2$ signals are found to have a
larger magnitude in \PbPb\ than in \pPb, while the $v_3$ signal has
a remarkably similar magnitude in both systems.
In addition, the long-range yield, $v_2$ obtained from the four-particle method, and $v_3$
all become apparent at about the same multiplicity.
The comprehensive correlation data presented in this paper, spanning a very wide range in particle
multiplicity and transverse momentum, should provide significant insights into
the origin of the azimuthal correlations in small collision systems,
particularly in the context of the hydrodynamic and color glass condensate models.

\section*{Acknowledgments}
We congratulate our colleagues in the CERN accelerator departments for the excellent performance of the LHC and thank the technical and administrative staffs at CERN and at other CMS institutes for their contributions to the success of the CMS effort. In addition, we gratefully acknowledge the computing centres and personnel of the Worldwide LHC Computing Grid for delivering so effectively the computing infrastructure essential to our analyses. Finally, we acknowledge the enduring support for the construction and operation of the LHC and the CMS detector provided by the following funding agencies: BMWF and FWF (Austria); FNRS and FWO (Belgium); CNPq, CAPES, FAPERJ, and FAPESP (Brazil); MEYS (Bulgaria); CERN; CAS, MoST, and NSFC (China); COLCIENCIAS (Colombia); MSES (Croatia); RPF (Cyprus); MoER, SF0690030s09 and ERDF (Estonia); Academy of Finland, MEC, and HIP (Finland); CEA and CNRS/IN2P3 (France); BMBF, DFG, and HGF (Germany); GSRT (Greece); OTKA and NKTH (Hungary); DAE and DST (India); IPM (Iran); SFI (Ireland); INFN (Italy); NRF and WCU (Republic of Korea); LAS (Lithuania); CINVESTAV, CONACYT, SEP, and UASLP-FAI (Mexico); MSI (New Zealand); PAEC (Pakistan); MSHE and NSC (Poland); FCT (Portugal); JINR (Armenia, Belarus, Georgia, Ukraine, Uzbekistan); MON, RosAtom, RAS and RFBR (Russia); MSTD (Serbia); SEIDI and CPAN (Spain); Swiss Funding Agencies (Switzerland); NSC (Taipei); ThEPCenter, IPST and NSTDA (Thailand); TUBITAK and TAEK (Turkey); NASU (Ukraine); STFC (United Kingdom); DOE and NSF (USA).

Individuals have received support from the Marie-Curie programme and the European Research Council and EPLANET (European Union); the Leventis Foundation; the A. P. Sloan Foundation; the Alexander von Humboldt Foundation; the Belgian Federal Science Policy Office; the Fonds pour la Formation \`a la Recherche dans l'Industrie et dans l'Agriculture (FRIA-Belgium); the Agentschap voor Innovatie door Wetenschap en Technologie (IWT-Belgium); the Ministry of Education, Youth and Sports (MEYS) of Czech Republic; the Council of Science and Industrial Research, India; the Compagnia di San Paolo (Torino); the HOMING PLUS programme of Foundation for Polish Science, cofinanced by EU, Regional Development Fund; and the Thalis and Aristeia programmes cofinanced by EU-ESF and the Greek NSRF.

\bibliography{auto_generated}   

\providecommand{\href}[2]{#2}\begingroup\raggedright\begin{thebibliography}{10}%
\makeatletter
\providecommand{\hrefCMSnoop }[0]{\@secondoftwo}%
\makeatother
\providecommand{\doi}{\texttt{doi:}\begingroup \urlstyle{tt}\Url}

\bibitem{Ollitrault:1992bk}
\hrefCMSnoop {} {J.-Y. Ollitrault, ``{Anisotropy as a signature of transverse
  collective flow}'',} \textit{ Phys. Rev. D} \textbf{ 46} (1992) 229,
\href{http://dx.doi.org/10.1103/PhysRevD.46.229}{\doi{10.1103/PhysRevD.46.229}}.

\bibitem{Alver:2008gk}
\hrefCMSnoop {} {{ PHOBOS} Collaboration, ``{System size dependence of cluster
  properties from two- particle angular correlations in Cu+Cu and Au+Au
  collisions at {\rootsNN} = 200\GeV}'',} \textit{ Phys. Rev. C} \textbf{ 81}
  (2010) 024904,
  \href{http://dx.doi.org/10.1103/PhysRevC.81.024904}{\doi{10.1103/PhysRevC.81.024904}},
\href{http://www.arXiv.org/abs/0812.1172}{\texttt{ arXiv:0812.1172}}.

\bibitem{Adcox:2004mh}
\hrefCMSnoop {} {{ PHENIX} Collaboration, ``{Formation of dense partonic matter
  in relativistic nucleus nucleus collisions at RHIC: Experimental evaluation
  by the PHENIX collaboration}'',} \textit{ Nucl. Phys. A} \textbf{ 757} (2005)
  184,
  \href{http://dx.doi.org/10.1016/j.nuclphysa.2005.03.086}{\doi{10.1016/j.nuclphysa.2005.03.086}},
\href{http://www.arXiv.org/abs/nucl-ex/0410003}{\texttt{
  arXiv:nucl-ex/0410003}}.

\bibitem{Adams:2005dq}
\hrefCMSnoop {} {{ STAR} Collaboration, ``{Experimental and theoretical
  challenges in the search for the quark gluon plasma: The STAR collaboration's
  critical assessment of the evidence from RHIC collisions}'',} \textit{ Nucl.
  Phys. A} \textbf{ 757} (2005) 102,
  \href{http://dx.doi.org/10.1016/j.nuclphysa.2005.03.085}{\doi{10.1016/j.nuclphysa.2005.03.085}},
\href{http://www.arXiv.org/abs/nucl-ex/0501009}{\texttt{
  arXiv:nucl-ex/0501009}}.

\bibitem{Back:2004je}
\hrefCMSnoop {} {B.~B. Back {et~al.}, ``{The PHOBOS perspective on discoveries
  at RHIC}'',} \textit{ Nucl. Phys. A} \textbf{ 757} (2005) 28,
  \href{http://dx.doi.org/10.1016/j.nuclphysa.2005.03.084}{\doi{10.1016/j.nuclphysa.2005.03.084}},
\href{http://www.arXiv.org/abs/nucl-ex/0410022}{\texttt{
  arXiv:nucl-ex/0410022}}.

\bibitem{Adler:2004cj}
\hrefCMSnoop {} {{ PHENIX} Collaboration, ``{Saturation of azimuthal anisotropy
  in Au + Au collisions at $\sqrt{s_{NN}} =$ 62 GeV to 200 GeV}'',} \textit{
  Phys. Rev. Lett.} \textbf{ 94} (2005) 232302,
\href{http://dx.doi.org/10.1103/PhysRevLett.94.232302}{\doi{10.1103/PhysRevLett.94.232302}}.

\bibitem{Agakishiev:2011id}
\hrefCMSnoop {} {{ STAR} Collaboration, ``{Directed and elliptic flow of
  charged particles in Cu+Cu collisions at $\sqrt{s_{NN}} =$ 22.4 GeV}'',}
  \textit{ Phys. Rev. C} \textbf{ 85} (2012) 014901,
\href{http://dx.doi.org/10.1103/PhysRevC.85.014901}{\doi{10.1103/PhysRevC.85.014901}}.

\bibitem{PHOBOS:PhysRevLett.98.242302}
\hrefCMSnoop {} {{ PHOBOS} Collaboration, ``System Size, Energy,
  Pseudorapidity, and Centrality Dependence of Elliptic Flow'',} \textit{ Phys.
  Rev. Lett.} \textbf{ 98} (2007) 242302,
  \href{http://dx.doi.org/10.1103/PhysRevLett.98.242302}{\doi{10.1103/PhysRevLett.98.242302}}.

\bibitem{Chatrchyan:2012ta}
\hrefCMSnoop {} {{ CMS} Collaboration, ``{Measurement of the elliptic
  anisotropy of charged particles produced in PbPb collisions at
  nucleon-nucleon center-of-mass energy = 2.76 TeV}'',} \textit{ Phys. Rev. C}
  \textbf{ 87} (2013) 014902,
  \href{http://dx.doi.org/10.1103/PhysRevC.87.014902}{\doi{10.1103/PhysRevC.87.014902}},
\href{http://www.arXiv.org/abs/1204.1409}{\texttt{ arXiv:1204.1409}}.

\bibitem{ALICE:PhysRevLett.105.252302}
\hrefCMSnoop {} {{ ALICE} Collaboration, ``Elliptic Flow of Charged Particles
  in Pb-Pb Collisions at {$\sqrt{s_{NN}}=2.76\TeV$}'',} \textit{ Phys. Rev.
  Lett.} \textbf{ 105} (2010) 252302,
  \href{http://dx.doi.org/10.1103/PhysRevLett.105.252302}{\doi{10.1103/PhysRevLett.105.252302}}.

\bibitem{ATLAS:2011ah}
\hrefCMSnoop {} {{ ATLAS} Collaboration, ``{Measurement of the pseudorapidity
  and transverse momentum dependence of the elliptic flow of charged particles
  in lead-lead collisions at {{\rootsNN} = 2.76\TeV} with the ATLAS
  detector}'',} \textit{ Phys. Lett. B} \textbf{ 707} (2012) 330,
  \href{http://dx.doi.org/10.1016/j.physletb.2011.12.056}{\doi{10.1016/j.physletb.2011.12.056}},
\href{http://www.arXiv.org/abs/1108.6018}{\texttt{ arXiv:1108.6018}}.

\bibitem{Adams:2005ph}
\hrefCMSnoop {} {{ STAR} Collaboration, ``{Distributions of charged hadrons
  associated with high transverse momentum particles in pp and Au + Au
  collisions at {\rootsNN} = 200\GeV}'',} \textit{ Phys. Rev. Lett.} \textbf{
  95} (2005) 152301,
  \href{http://dx.doi.org/10.1103/PhysRevLett.95.152301}{\doi{10.1103/PhysRevLett.95.152301}},
\href{http://www.arXiv.org/abs/nucl-ex/0501016}{\texttt{
  arXiv:nucl-ex/0501016}}.

\bibitem{Abelev:2009af}
\hrefCMSnoop {} {{ STAR} Collaboration, ``{Long range rapidity correlations and
  jet production in high energy nuclear collisions}'',} \textit{ Phys. Rev. C}
  \textbf{ 80} (2009) 064912,
  \href{http://dx.doi.org/10.1103/PhysRevC.80.064912}{\doi{10.1103/PhysRevC.80.064912}},
\href{http://www.arXiv.org/abs/0909.0191}{\texttt{ arXiv:0909.0191}}.

\bibitem{Alver:2009id}
\hrefCMSnoop {} {{ PHOBOS} Collaboration, ``{High transverse momentum triggered
  correlations over a large pseudorapidity acceptance in Au+Au collisions at
  {\rootsNN} = 200\GeV}'',} \textit{ Phys. Rev. Lett.} \textbf{ 104} (2010)
  062301,
  \href{http://dx.doi.org/10.1103/PhysRevLett.104.062301}{\doi{10.1103/PhysRevLett.104.062301}},
\href{http://www.arXiv.org/abs/0903.2811}{\texttt{ arXiv:0903.2811}}.

\bibitem{Abelev:2009jv}
\hrefCMSnoop {} {{ STAR} Collaboration, ``{Three-particle coincidence of the
  long range pseudorapidity correlation in high energy nucleus-nucleus
  collisions}'',} \textit{ Phys. Rev. Lett.} \textbf{ 105} (2010) 022301,
  \href{http://dx.doi.org/10.1103/PhysRevLett.105.022301}{\doi{10.1103/PhysRevLett.105.022301}},
\href{http://www.arXiv.org/abs/0912.3977}{\texttt{ arXiv:0912.3977}}.

\bibitem{Armesto:2004pt}
\hrefCMSnoop {} {N.~Armesto, C.~A. Salgado, and U.~A. Wiedemann, ``{Measuring
  the collective flow with jets}'',} \textit{ Phys. Rev. Lett.} \textbf{ 93}
  (2004) 242301,
  \href{http://dx.doi.org/10.1103/PhysRevLett.93.242301}{\doi{10.1103/PhysRevLett.93.242301}},
\href{http://www.arXiv.org/abs/hep-ph/0405301}{\texttt{ arXiv:hep-ph/0405301}}.

\bibitem{Majumder:2006wi}
\hrefCMSnoop {} {A.~Majumder, B.~Muller, and S.~A. Bass, ``{Longitudinal
  Broadening of Quenched Jets in Turbulent Color Fields}'',} \textit{ Phys.
  Rev. Lett.} \textbf{ 99} (2007) 042301,
  \href{http://dx.doi.org/10.1103/PhysRevLett.99.042301}{\doi{10.1103/PhysRevLett.99.042301}},
\href{http://www.arXiv.org/abs/hep-ph/0611135}{\texttt{ arXiv:hep-ph/0611135}}.

\bibitem{Chiu:2005ad}
\hrefCMSnoop {} {C.~B. Chiu and R.~C. Hwa, ``{Pedestal and peak structure in
  jet correlation}'',} \textit{ Phys. Rev. C} \textbf{ 72} (2005) 034903,
  \href{http://dx.doi.org/10.1103/PhysRevC.72.034903}{\doi{10.1103/PhysRevC.72.034903}},
\href{http://www.arXiv.org/abs/nucl-th/0505014}{\texttt{
  arXiv:nucl-th/0505014}}.

\bibitem{Wong:2008yh}
\hrefCMSnoop {} {C.-Y. Wong, ``Momentum kick model description of the near-side
  ridge and jet quenching'',} \textit{ Phys. Rev. C} \textbf{ 78} (2008)
  064905,
  \href{http://dx.doi.org/10.1103/PhysRevC.78.064905}{\doi{10.1103/PhysRevC.78.064905}},
\href{http://www.arXiv.org/abs/0806.2154}{\texttt{ arXiv:0806.2154}}.

\bibitem{Romatschke:2006bb}
\hrefCMSnoop {} {P.~Romatschke, ``{Momentum broadening in an anisotropic
  plasma}'',} \textit{ Phys. Rev. C} \textbf{ 75} (2007) 014901,
  \href{http://dx.doi.org/10.1103/PhysRevC.75.014901}{\doi{10.1103/PhysRevC.75.014901}},
\href{http://www.arXiv.org/abs/hep-ph/0607327}{\texttt{ arXiv:hep-ph/0607327}}.

\bibitem{Shuryak:2007fu}
\hrefCMSnoop {} {E.~V. Shuryak, ``On the origin of the ``ridge'' phenomenon
  induced by jets in heavy ion collisions'',} \textit{ Phys. Rev. C} \textbf{
  76} (2007) 047901,
  \href{http://dx.doi.org/10.1103/PhysRevC.76.047901}{\doi{10.1103/PhysRevC.76.047901}},
\href{http://www.arXiv.org/abs/0706.3531}{\texttt{ arXiv:0706.3531}}.

\bibitem{Voloshin:2004th}
\hrefCMSnoop {} {S.~A. Voloshin, ``{Two particle rapidity, transverse momentum,
  and azimuthal correlations in relativistic nuclear collisions and transverse
  radial expansion}'',} \textit{ Nucl. Phys. A} \textbf{ 749} (2005) 287,
  \href{http://dx.doi.org/10.1016/j.nuclphysa.2004.12.053}{\doi{10.1016/j.nuclphysa.2004.12.053}},
\href{http://www.arXiv.org/abs/nucl-th/0410024}{\texttt{
  arXiv:nucl-th/0410024}}.

\bibitem{Mishra:2007tw}
\hrefCMSnoop {} {A.~P. Mishra, R.~K. Mohapatra, P.~S. Saumia, and A.~M.
  Srivastava, ``{Superhorizon fluctuations and acoustic oscillations in
  relativistic heavy-ion collisions}'',} \textit{ Phys. Rev. C} \textbf{ 77}
  (2008) 064902,
  \href{http://dx.doi.org/10.1103/PhysRevC.77.064902}{\doi{10.1103/PhysRevC.77.064902}},
\href{http://www.arXiv.org/abs/0711.1323}{\texttt{ arXiv:0711.1323}}.

\bibitem{Takahashi:2009na}
\hrefCMSnoop {} {J.~Takahashi {et~al.}, ``{Topology studies of hydrodynamics
  using two particle correlation analysis}'',} \textit{ Phys. Rev. Lett.}
  \textbf{ 103} (2009) 242301,
  \href{http://dx.doi.org/10.1103/PhysRevLett.103.242301}{\doi{10.1103/PhysRevLett.103.242301}},
\href{http://www.arXiv.org/abs/0902.4870}{\texttt{ arXiv:0902.4870}}.

\bibitem{Alver:2010gr}
\hrefCMSnoop {} {B.~Alver and G.~Roland, ``{Collision geometry fluctuations and
  triangular flow in heavy-ion collisions}'',} \textit{ Phys. Rev. C} \textbf{
  81} (2010) 054905,
  \href{http://dx.doi.org/10.1103/PhysRevC.81.054905}{\doi{10.1103/PhysRevC.81.054905}},
  \href{http://www.arXiv.org/abs/1003.0194}{\texttt{ arXiv:1003.0194}}.
Erratum: \doi{10.1103/PhysRevC.82.039903}.

\bibitem{Alver:2010dn}
\hrefCMSnoop {} {B.~H. Alver, C.~Gombeaud, M.~Luzum, and J.-Y. Ollitrault,
  ``{Triangular flow in hydrodynamics and transport theory}'',} \textit{ Phys.
  Rev. C} \textbf{ 82} (2010) 034913,
  \href{http://dx.doi.org/10.1103/PhysRevC.82.034913}{\doi{10.1103/PhysRevC.82.034913}},
\href{http://www.arXiv.org/abs/1007.5469}{\texttt{ arXiv:1007.5469}}.

\bibitem{Schenke:2010rr}
\hrefCMSnoop {} {B.~Schenke, S.~Jeon, and C.~Gale, ``Elliptic and triangular
  flow in event-by-event {D=3+1} viscous hydrodynamics'',} \textit{ Phys. Rev.
  Lett.} \textbf{ 106} (2011) 042301,
  \href{http://dx.doi.org/10.1103/PhysRevLett.106.042301}{\doi{10.1103/PhysRevLett.106.042301}},
\href{http://www.arXiv.org/abs/1009.3244}{\texttt{ arXiv:1009.3244}}.

\bibitem{Petersen:2010cw}
\hrefCMSnoop {} {H.~Petersen, G.-Y. Qin, S.~A. Bass, and B.~M{\"u}ller,
  ``{Triangular flow in event-by-event ideal hydrodynamics in Au+Au collisions
  at {\rootsNN} = 200 GeV}'',} \textit{ Phys. Rev. C} \textbf{ 82} (2010)
  041901,
  \href{http://dx.doi.org/10.1103/PhysRevC.82.041901}{\doi{10.1103/PhysRevC.82.041901}},
\href{http://www.arXiv.org/abs/1008.0625}{\texttt{ arXiv:1008.0625}}.

\bibitem{Xu:2010du}
\hrefCMSnoop {} {J.~Xu and C.~M. Ko, ``Effects of triangular flow on di-hadron
  azimuthal correlations in relativistic heavy ion collisions'',} \textit{
  Phys. Rev. C} \textbf{ 83} (2011) 021903,
  \href{http://dx.doi.org/10.1103/PhysRevC.83.021903}{\doi{10.1103/PhysRevC.83.021903}},
\href{http://www.arXiv.org/abs/1011.3750}{\texttt{ arXiv:1011.3750}}.

\bibitem{Teaney:2010vd}
\hrefCMSnoop {} {D.~Teaney and L.~Yan, ``Triangularity and dipole asymmetry in
  heavy ion collisions'',} \textit{ Phys. Rev. C} \textbf{ 83} (2011) 064904,
  \href{http://dx.doi.org/10.1103/PhysRevC.83.064904}{\doi{10.1103/PhysRevC.83.064904}},
\href{http://www.arXiv.org/abs/1010.1876}{\texttt{ arXiv:1010.1876}}.

\bibitem{Voloshin:1994mz}
\hrefCMSnoop {} {S.~Voloshin and Y.~Zhang, ``{Flow study in relativistic
  nuclear collisions by Fourier expansion of azimuthal particle
  distributions}'',} \textit{ Z. Phys. C} \textbf{ 70} (1996) 665,
  \href{http://dx.doi.org/10.1007/s002880050141}{\doi{10.1007/s002880050141}},
\href{http://www.arXiv.org/abs/hep-ph/9407282}{\texttt{ arXiv:hep-ph/9407282}}.

\bibitem{Qiu:2011hf}
\hrefCMSnoop {} {Z.~Qiu, C.~Shen, and U.~Heinz, ``{Hydrodynamic elliptic and
  triangular flow in Pb-Pb collisions at {\rootsNN} = 2.76\TeV}'',} \textit{
  Phys. Lett. B} \textbf{ 707} (2012) 151,
  \href{http://dx.doi.org/10.1016/j.physletb.2011.12.041}{\doi{10.1016/j.physletb.2011.12.041}},
\href{http://www.arXiv.org/abs/1110.3033}{\texttt{ arXiv:1110.3033}}.

\bibitem{Chatrchyan:2011eka}
\hrefCMSnoop {} {{ CMS} Collaboration, ``{Long-range and short-range dihadron
  angular correlations in central PbPb collisions at a nucleon-nucleon center
  of mass energy of 2.76 TeV}'',} \textit{ JHEP} \textbf{ 07} (2011) 076,
  \href{http://dx.doi.org/10.1007/JHEP07(2011)076}{\doi{10.1007/JHEP07(2011)076}},
\href{http://www.arXiv.org/abs/1105.2438}{\texttt{ arXiv:1105.2438}}.

\bibitem{Chatrchyan:2012wg}
\hrefCMSnoop {} {{ CMS} Collaboration, ``{Centrality dependence of dihadron
  correlations and azimuthal anisotropy harmonics in PbPb collisions at
  {\rootsNN} = 2.76 TeV}'',} \textit{ Eur. Phys. J. C} \textbf{ 72} (2012)
  2012,
  \href{http://dx.doi.org/10.1140/epjc/s10052-012-2012-3}{\doi{10.1140/epjc/s10052-012-2012-3}},
\href{http://www.arXiv.org/abs/1201.3158}{\texttt{ arXiv:1201.3158}}.

\bibitem{ALICE:2011ab}
\hrefCMSnoop {} {{ ALICE} Collaboration, ``{Higher harmonic anisotropic flow
  measurements of charged particles in Pb-Pb collisions at {\rootsNN} =
  2.76\TeV }'',} \textit{ Phys. Rev. Lett.} \textbf{ 107} (2011) 032301,
  \href{http://dx.doi.org/10.1103/PhysRevLett.107.032301}{\doi{10.1103/PhysRevLett.107.032301}},
\href{http://www.arXiv.org/abs/1105.3865}{\texttt{ arXiv:1105.3865}}.

\bibitem{Aamodt:2011by}
\hrefCMSnoop {} {{ ALICE} Collaboration, ``{Harmonic decomposition of
  two-particle angular correlations in Pb-Pb collisions at {\rootsNN} =
  2.76\TeV}'',} \textit{ Phys. Lett. B} \textbf{ 708} (2012) 249,
  \href{http://dx.doi.org/10.1016/j.physletb.2012.01.060}{\doi{10.1016/j.physletb.2012.01.060}},
\href{http://www.arXiv.org/abs/1109.2501}{\texttt{ arXiv:1109.2501}}.

\bibitem{ATLAS:2012at}
\hrefCMSnoop {} {{ ATLAS} Collaboration, ``{Measurement of the azimuthal
  anisotropy for charged particle production in {\rootsNN} = 2.76\TeV lead-lead
  collisions with the ATLAS detector}'',} \textit{ Phys. Rev. C} \textbf{ 86}
  (2012) 014907,
  \href{http://dx.doi.org/10.1103/PhysRevC.86.014907}{\doi{10.1103/PhysRevC.86.014907}},
\href{http://www.arXiv.org/abs/1203.3087}{\texttt{ arXiv:1203.3087}}.

\bibitem{Khachatryan:2010gv}
\hrefCMSnoop {} {{ CMS} Collaboration, ``{Observation of long-range near-side
  angular correlations in proton-proton collisions at the LHC}'',} \textit{
  JHEP} \textbf{ 09} (2010) 091,
  \href{http://dx.doi.org/10.1007/JHEP09(2010)091}{\doi{10.1007/JHEP09(2010)091}},
\href{http://www.arXiv.org/abs/1009.4122}{\texttt{ arXiv:1009.4122}}.

\bibitem{CMS:2012qk}
\hrefCMSnoop {} {{ CMS} Collaboration, ``{Observation of long-range near-side
  angular correlations in proton-lead collisions at the LHC}'',} \textit{ Phys.
  Lett. B} \textbf{ 718} (2013) 795,
  \href{http://dx.doi.org/10.1016/j.physletb.2012.11.025}{\doi{10.1016/j.physletb.2012.11.025}},
\href{http://www.arXiv.org/abs/1210.5482}{\texttt{ arXiv:1210.5482}}.

\bibitem{alice:2012qe}
\hrefCMSnoop {} {{ ALICE} Collaboration, ``{Long-range angular correlations on
  the near and away side in \pPb\ collisions at {\rootsNN} = 5.02\TeV }'',}
  \textit{ Phys. Lett. B} \textbf{ 719} (2013) 29,
  \href{http://dx.doi.org/10.1016/j.physletb.2013.01.012}{\doi{10.1016/j.physletb.2013.01.012}},
\href{http://www.arXiv.org/abs/1212.2001}{\texttt{ arXiv:1212.2001}}.

\bibitem{atlas:2012fa}
\hrefCMSnoop {} {{ ATLAS} Collaboration, ``{Observation of Associated Near-Side
  and Away-Side Long-Range Correlations in {\rootsNN} = 5.02\TeV Proton-lead
  Collisions with the ATLAS Detector}'',} \textit{ Phys. Rev. Lett.} \textbf{
  110} (2013) 182302,
  \href{http://dx.doi.org/10.1103/PhysRevLett.110.182302}{\doi{10.1103/PhysRevLett.110.182302}},
\href{http://www.arXiv.org/abs/1212.5198}{\texttt{ arXiv:1212.5198}}.

\bibitem{Adare:2013piz}
\hrefCMSnoop {} {{ PHENIX} Collaboration, ``{Quadrupole anisotropy in dihadron
  azimuthal correlations in central d+Au collisions at {\rootsNN} =
  200~GeV}'',} (2013). \href{http://www.arXiv.org/abs/1303.1794}{\texttt{
  arXiv:1303.1794}}.
Submitted to Phys. Rev. Lett.

\bibitem{Li:2012hc}
\hrefCMSnoop {} {W.~Li, ``{Observation of a `Ridge' correlation structure in
  high multiplicity proton-proton collisions: A brief review}'',} \textit{ Mod.
  Phys. Lett. A} \textbf{ 27} (2012) 1230018,
  \href{http://dx.doi.org/10.1142/S0217732312300182}{\doi{10.1142/S0217732312300182}},
\href{http://www.arXiv.org/abs/1206.0148}{\texttt{ arXiv:1206.0148}}.

\bibitem{Dusling:2012cg}
\hrefCMSnoop {} {K.~Dusling and R.~Venugopalan, ``Evidence for {BFKL} and
  saturation dynamics from dihadron spectra at the {LHC}'',} \textit{ Phys.
  Rev. D} \textbf{ 87} (2013) 051502,
  \href{http://dx.doi.org/10.1103/PhysRevD.87.051502}{\doi{10.1103/PhysRevD.87.051502}},
\href{http://www.arXiv.org/abs/1210.3890}{\texttt{ arXiv:1210.3890}}.

\bibitem{Dusling:2012wy}
\hrefCMSnoop {} {K.~Dusling and R.~Venugopalan, ``Explanation of systematics of
  {CMS p+Pb} high multiplicity dihadron data at {\rootsNN} = 5.02 {TeV}'',}
  \textit{ Phys. Rev. D} \textbf{ 87} (2013) 054014,
  \href{http://dx.doi.org/10.1103/PhysRevD.87.054014}{\doi{10.1103/PhysRevD.87.054014}},
\href{http://www.arXiv.org/abs/1211.3701}{\texttt{ arXiv:1211.3701}}.

\bibitem{Bozek:2011if}
\hrefCMSnoop {} {P.~Bozek, ``{Collective flow in p-Pb and d-Pd collisions at
  TeV energies}'',} \textit{ Phys. Rev. C} \textbf{ 85} (2012) 014911,
  \href{http://dx.doi.org/10.1103/PhysRevC.85.014911}{\doi{10.1103/PhysRevC.85.014911}},
\href{http://www.arXiv.org/abs/1112.0915}{\texttt{ arXiv:1112.0915}}.

\bibitem{Bozek:2012gr}
\hrefCMSnoop {} {P.~Bozek and W.~Broniowski, ``{Correlations from hydrodynamic
  flow in \pPb\ collisions}'',} \textit{ Phys. Lett. B} \textbf{ 718} (2013)
  1557,
  \href{http://dx.doi.org/10.1016/j.physletb.2012.12.051}{\doi{10.1016/j.physletb.2012.12.051}},
\href{http://www.arXiv.org/abs/1211.0845}{\texttt{ arXiv:1211.0845}}.

\bibitem{Bilandzic:2010jr}
\hrefCMSnoop {} {A.~Bilandzic, R.~Snellings, and S.~Voloshin, ``{Flow analysis
  with cumulants: Direct calculations}'',} \textit{ Phys. Rev. C} \textbf{ 83}
  (2011) 044913,
  \href{http://dx.doi.org/10.1103/PhysRevC.83.044913}{\doi{10.1103/PhysRevC.83.044913}},
\href{http://www.arXiv.org/abs/1010.0233}{\texttt{ arXiv:1010.0233}}.

\bibitem{Aad:2013fja}
\hrefCMSnoop {} {{ ATLAS} Collaboration, ``{Measurement with the ATLAS detector
  of multi-particle azimuthal correlations in p+Pb collisions at {\rootsNN} =
  5.02~TeV}'',} (2013). \href{http://www.arXiv.org/abs/1303.2084}{\texttt{
  arXiv:1303.2084}}.
Submitted to Phys. Lett. B.

\bibitem{JINST}
\hrefCMSnoop {} {{ CMS} Collaboration, ``The {CMS} experiment at the {CERN}
  {LHC}'',} \textit{ JINST} \textbf{ 3} (2008) S08004,
\href{http://dx.doi.org/10.1088/1748-0221/3/08/S08004}{\doi{10.1088/1748-0221/3/08/S08004}}.

\bibitem{GEANT4}
\hrefCMSnoop {} {{ Geant4} Collaboration, ``{Geant4---a simulation toolkit}'',}
  \textit{ Nucl. Instrum. and Meth. A} \textbf{ 506} (2003) 250,
\href{http://dx.doi.org/10.1016/S0168-9002(03)01368-8}{\doi{10.1016/S0168-9002(03)01368-8}}.

\bibitem{Porteboeuf:2010um}
\hrefCMSnoop {} {S.~Porteboeuf, T.~Pierog, and K.~Werner, ``{Producing Hard
  Processes Regarding the Complete Event: The EPOS Event Generator}'',} (2010).
\href{http://www.arXiv.org/abs/1006.2967}{\texttt{ arXiv:1006.2967}}.

\bibitem{Gyulassy:1994ew}
\hrefCMSnoop {} {M.~Gyulassy and X.-N. Wang, ``{HIJING 1.0: A Monte Carlo
  program for parton and particle production in high-energy hadronic and
  nuclear collisions}'',} \textit{ Comput. Phys. Commun.} \textbf{ 83} (1994)
  307,
  \href{http://dx.doi.org/10.1016/0010-4655(94)90057-4}{\doi{10.1016/0010-4655(94)90057-4}},
\href{http://www.arXiv.org/abs/nucl-th/9502021}{\texttt{
  arXiv:nucl-th/9502021}}.

\bibitem{Stenson:2010xx}
\href {http://cdsweb.cern.ch/record/1258204} {{ CMS} Collaboration, ``Tracking
  and Vertexing Results from First Collisions'',} CMS Physics Analysis Summary
  CMS-PAS-TRK-10-001, (2010).

\bibitem{Chatrchyan:2012xq}
\hrefCMSnoop {} {{ CMS} Collaboration, ``{Azimuthal anisotropy of charged
  particles at high transverse momenta in PbPb collisions at {\rootsNN} =
  2.76\TeV}'',} \textit{ Phys. Rev. Lett.} \textbf{ 109} (2012) 022301,
  \href{http://dx.doi.org/10.1103/PhysRevLett.109.022301}{\doi{10.1103/PhysRevLett.109.022301}},
\href{http://www.arXiv.org/abs/1204.1850}{\texttt{ arXiv:1204.1850}}.

\bibitem{TRK-10-002}
\href {http://cdsweb.cern.ch/record/1279139} {{ CMS} Collaboration,
  ``Measurement of Tracking Efficiency'',} CMS Physics Analysis Summary
  CMS-PAS-TRK-10-002, (2010).

\bibitem{PhysRevC.72.011902}
N.~N. Ajitanand\hrefCMSnoop {} { {et~al.}, ``{Decomposition of harmonic and jet
  contributions to particle-pair correlations at ultrarelativistic
  energies}'',} \textit{ Phys. Rev. C} \textbf{ 72} (2005) 011902,
  \href{http://dx.doi.org/10.1103/PhysRevC.72.011902}{\doi{10.1103/PhysRevC.72.011902}},
\href{http://www.arXiv.org/abs/nucl-ex/0501025}{\texttt{
  arXiv:nucl-ex/0501025}}.

\bibitem{Dusling:2013}
\hrefCMSnoop {} {K.~Dusling and R.~Venugopalan, ``{Comparison of the Color
  Glass Condensate to di-hadron correlations in proton-proton and
  proton-nucleus collisions}'',} (2013).
\href{http://www.arXiv.org/abs/1302.7018}{\texttt{ arXiv:1302.7018}}.

\bibitem{Ollitrault:2009ie}
\hrefCMSnoop {} {J.-Y. Ollitrault, A.~M. Poskanzer, and S.~A. Voloshin,
  ``{Effect of flow fluctuations and nonflow on elliptic flow methods}'',}
  \textit{ Phys. Rev. C} \textbf{ 80} (2009) 014904,
  \href{http://dx.doi.org/10.1103/PhysRevC.80.014904}{\doi{10.1103/PhysRevC.80.014904}},
\href{http://www.arXiv.org/abs/0904.2315}{\texttt{ arXiv:0904.2315}}.

\bibitem{Alver:2007qw}
\hrefCMSnoop {} {{ PHOBOS} Collaboration, ``{Event-by-Event Fluctuations of
  Azimuthal Particle Anisotropy in Au + Au Collisions at {\rootsNN} =
  200\GeV}'',} \textit{ Phys. Rev. Lett.} \textbf{ 104} (2010) 142301,
  \href{http://dx.doi.org/10.1103/PhysRevLett.104.142301}{\doi{10.1103/PhysRevLett.104.142301}},
\href{http://www.arXiv.org/abs/nucl-ex/0702036}{\texttt{
  arXiv:nucl-ex/0702036}}.

\end{thebibliography}\endgroup

\cleardoublepage \appendix\section{The CMS Collaboration \label{app:collab}}\begin{sloppypar}\hyphenpenalty=5000\widowpenalty=500\clubpenalty=5000\textbf{Yerevan Physics Institute,  Yerevan,  Armenia}\\*[0pt]
S.~Chatrchyan, V.~Khachatryan, A.M.~Sirunyan, A.~Tumasyan
\vskip\cmsinstskip
\textbf{Institut f\"{u}r Hochenergiephysik der OeAW,  Wien,  Austria}\\*[0pt]
W.~Adam, T.~Bergauer, M.~Dragicevic, J.~Er\"{o}, C.~Fabjan\cmsAuthorMark{1}, M.~Friedl, R.~Fr\"{u}hwirth\cmsAuthorMark{1}, V.M.~Ghete, N.~H\"{o}rmann, J.~Hrubec, M.~Jeitler\cmsAuthorMark{1}, W.~Kiesenhofer, V.~Kn\"{u}nz, M.~Krammer\cmsAuthorMark{1}, I.~Kr\"{a}tschmer, D.~Liko, I.~Mikulec, D.~Rabady\cmsAuthorMark{2}, B.~Rahbaran, C.~Rohringer, H.~Rohringer, R.~Sch\"{o}fbeck, J.~Strauss, A.~Taurok, W.~Treberer-Treberspurg, W.~Waltenberger, C.-E.~Wulz\cmsAuthorMark{1}
% ...
\vskip\cmsinstskip
% ...
1:~~Also at Vienna University of Technology, Vienna, Austria\\
2:~~Also at CERN, European Organization for Nuclear Research, Geneva, Switzerland\\
% ...

\end{sloppypar}
\end{document}